\documentclass[a4paper,fleqn,usenatbib]{mnras}

\usepackage[T1]{fontenc}
\usepackage{ae,aecompl}

\usepackage{graphicx}	
\usepackage{amsmath}	
\usepackage{amssymb}	

\usepackage[english]{babel}             
\usepackage[binary-units]{siunitx}
\usepackage{xcolor,cancel}
\usepackage{psfrag}
\usepackage{framed}
\usepackage{graphicx-psmin}
\usepackage{pstricks,pst-node,pst-text,pst-3d,pst-plot,pst-3dplot,pstricks-add,pst-vue3d}
\usepackage{wrapfig}
\usepackage{multirow}
\usepackage[normalem]{ulem}
\usepackage{xspace} 

\newcommand{\sign}[1]{\text{sign}\left(#1\right)}
\newcommand{\td}[2]{\frac{\mathrm{d}{#1}}{\mathrm{d}{#2}}}
\newcommand{\ignore}[1]{}%
\newcommand{\vect}[1]{\boldsymbol{#1}}
\newcommand{\smat}[1]{\textbf{\textsf{#1}}}

\psset{linecolor=black} 
\newrgbcolor{darkgreen}{0 0.7 0.1}
\newrgbcolor{darkblue}{0 0.1 0.5}
\newrgbcolor{emphblue2}{0 0 1}
\newrgbcolor{emphyellow}{0.9 0.4 0.0}
\newrgbcolor{marron}{0.8 0.2 0.2}
\newrgbcolor{granate}{0.9 0.2 0.5}
\newrgbcolor{naranja}{0.9 0.4 0.0}
\newrgbcolor{verdeclaro}{0.7 0.99 0.8}
\newrgbcolor{azulclaro}{0.8 0.9 0.99}
\newrgbcolor{rosa}{0.99 0.9 0.9}
\newrgbcolor{Paleyellow}{1 1 0.78}
\newrgbcolor{ejes}{0 0 0} 

\newcommand{\sdg}{SDG\xspace} 

\title[On the MOID computation: a new effective method]{On the minimum orbital intersection distance computation: a new effective method}

\author[J. M. Hedo et al.]{
Jos\'e M. Hedo,$^{1}$\thanks{E-mail: josemanuel.hedo@upm.es}
Manuel Ru\'iz,$^{1}$
and Jes\'us Pel\'aez$^{1}$
\\
$^{1}$Space Dynamics Group, ETSIAE, Technical University of Madrid (UPM), Pza. Cardenal Cisneros, 3, 28040 Madrid, Spain\\
}

\date{Accepted XXX. Received YYY; in original form ZZZ}

\pubyear{2018}

\begin{document}
\label{firstpage}
\pagerange{\pageref{firstpage}--\pageref{lastpage}}
\maketitle

\begin{abstract}
The computation of the Minimum Orbital Intersection Distance (MOID) is an old,  but increasingly relevant problem. Fast and precise methods for MOID computation are needed to select potentially hazardous asteroids from a large catalogue. The same applies to debris with respect to spacecraft. An iterative method that strictly meets these two premises is presented.
\end{abstract}


\section{Introduction}

\begin{figure}
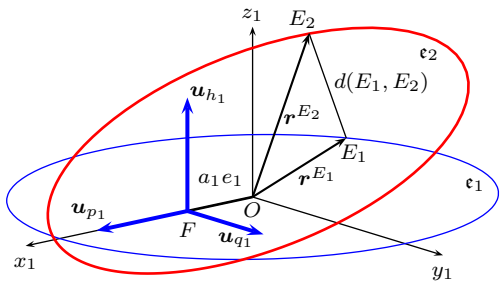

	\vspace{5mm}
	\centering
	\psset{unit=0.044\textwidth}%
	\psset{Alpha=40,Beta=15,RotSequence=zxy}%
	\pspicture(-4,-1)(5,4)%
	\small%
	
	\rput[tl](0.5,1){
		
		
		\pstThreeDPut(0,0,0){\rput[t](0.0,-0.1){$O$}}%
		\pstThreeDLine[linewidth=.5pt,linecolor=black,arrows=->](0,0,0)(5,0,0)%
		\pstThreeDPut(5,0,0){\rput[t](0,-.2){$x_1$}}%
		\pstThreeDLine[linewidth=.5pt,linecolor=black,arrows=->](0,0,0)(0,5,0)%
		\pstThreeDPut(0,5,0){\rput[t](0,-.2){$y_1$}}%
		\pstThreeDLine[linewidth=.5pt,linecolor=black,arrows=->](0,0,0)(0,0,3)%
		\pstThreeDPut(0,0,3){\rput[t](0,0.3){$z_1$}}%
		
		\pstThreeDEllipse[linewidth=0.5pt,linecolor=blue](0,0,0)(4.25,0,0)(0,4,0)%
		\pstThreeDLine[linewidth=1pt,linecolor=black](0,0,0)(1.436,0,0)%
		\pstThreeDPut(0.7,0,0){\rput[b](0,0.3){\textcolor{black}{$a_1 e_1$}}}%
		\pstThreeDPut(-3,2,0){\rput[bl](0,0){\textcolor{black}{$\mathfrak{e}_1$}}}%
		
		\pstThreeDPut(1.436,0,0) {
			\pstThreeDPut(0,0,0){\rput[t](0,-0.2){$F$}}%
			\pstThreeDLine[linewidth=1.5pt,linecolor=blue,arrows=->](0,0,0)(2,0,0)%
			\pstThreeDLine[linewidth=1.5pt,linecolor=blue,arrows=->](0,0,0)(0,2,0)%
			\pstThreeDLine[linewidth=1.5pt,linecolor=blue,arrows=->](0,0,0)(0,0,2)%
			\pstThreeDPut(2,0,0){\rput[br](0.2,0.2){\textcolor{black}{$\vect{u}_{p_1}$}}}%
			\pstThreeDPut(1,2,0){\rput[tl](0,.2){\textcolor{black}{$\vect{u}_{q_1}$}}}%
			\pstThreeDPut(0,0,2){\rput[br](0.7,-.1){\textcolor{black}{$\vect{u}_{h_1}$}}}%
			\pstThreeDPut(-7.2,-2.5,.9){\rput[bl](0,0){\textcolor{black}{$\mathfrak{e}_2$}}}
			\pstThreeDPut[RotZ=10,RotY=20](0,0,0){
				\parametricplotThreeD[linewidth=1.pt,linecolor=red,xPlotpoints=100](0,360){%
					3. t cos .5 mul 1. add div t cos mul 3. t cos .5 mul 1. add div t sin  mul 0. %
				}%
			}
		}
		
		\pstThreeDPut(-3,-1.8,2.3){\rput[b](-0.3,-0.2){\textcolor{black}{$E_2$}}}%
		\pstThreeDLine[arrows=->](0,0,0)(-2,-0.9,2.35)%
		\pstThreeDPut(0,0,0){\rput[b](0.85,1.25){\textcolor{black}{$\vect{r}^{E_2}$}}}%
		\pstThreeDPut(-4.2,-2.5,0){\rput[b](0.1,-0.6){\textcolor{black}{$E_1$}}}
		\pstThreeDLine[linecolor=black,arrows=->](0,0,0)(-3.7,-1.95,0)
		\pstThreeDPut(0,0,0){\rput[b](1.1,0.2){\textcolor{black}{$\vect{r}^{E_1}$}}}
		\pstThreeDLine[linewidth=0.5pt,linecolor=black,arrows=-](-3.7,-1.95,0)(-2,-0.9,2.35)%
		\pstThreeDPut(-2.8,-1.2,1.2){\rput[bl](0,-.1){\textcolor{black}{$d(E_1,E_2)$}}}%
	}
	\endpspicture
	\caption{Relative geometry of two elliptical confocal orbits.}
	\label{fig:TCE}
\end{figure}

The MOID is the distance between the closest points of the osculating orbits of two bodies. It can be computed as the minimum distance between two sets of points embedded in the Euclidean 3-space $(\mathbb{R}^3)$ -- see Fig.~\ref{fig:TCE}. One of the orbits is considered to be the reference or target (primary). Any other orbit facing the primary is called secondary, although the roles can be exchanged as needed.

MOID computation is a classical but  ever relevant problem. The risk of collision of a Potentially Hazardous Object (PHO) such as Apophis with the Earth \citep{wlodarczyk2013potentially} and of space debris  with an operational spacecraft \citep{casanova2014space} are two important applications.

Since the sets of secondaries are usually large, MOID and its time derivative can be used as a prefilter -- for example, the second level filter proposed by Hoots \citep{hoots1984analytic} --  to discard those that will not give problems in the immediate future. Those selected can be studied with more sophisticated methods -- see, for example, \citet{Bonnano}.

In most MOID calculations, orbits can be considered as  Keplerian ellipses. Some bodies originate outside of the Solar System. Recently the interstellar object {1I/2017 U1}, later named \textit{Oumuamua},  reached its perihelion (\SI{0.2483}{\astronomicalunit}) in a hyperbolic trajectory. In this case, MOID calculation with Earth involves a hyperbola. Thus, the  main task is to compute the minimum distance between two confocal Keplerian orbits, the primary -- Earth or spacecraft -- and one or more secondaries -- near-Earth objects or space debris.

Current methods for MOID computation are sufficient for identifying NEOs as Potentially Hazardous Asteroids, or space debris coming close to a spacecraft orbit. Faster algorithms for MOID computation could help to identify NEOs coming close to perturbing planets \citep{Sitarski}, which in time could change their orbits significantly; study evolution of perturbed NEO's MOID; manage growing catalogues of space debris; and compute covariance of NEO's and debris' MOID.

There is an abundant literature on MOID computation from the mid-20th century to the present day. Some authors use an algebraic approach, obtaining all critical points of the distance function. Others use numerical methods, finding the global minimum by iterations. Others use a hybrid approach, such as \citet{Derevyanka}. Some key results are summarized in Table~\ref{tab:Methods}.

\begin{table*}
	\caption{Timetable of some MOID computation methods.}
	\label{tab:Methods}
	\centering
	\begin{tabular}{l m{4.2cm} m{7cm}}
		\hline
		\multicolumn{1}{c}{\bf Author} & \multicolumn{1}{c}{\bf Application} & \multicolumn{1}{c}{\bf MOID computation method} \\ \hline
		
		\multicolumn{3}{c}{Algebraic (exact approach)}\\ \hline
		
		\citet{Dybczinski}  & Determination of the closest approach between two Keplerian orbits. & 		Solve the classical trigonometrical equations that provide the critical points of distance and whose variables are the orbit anomalies  (Sitarski, etc.) with procedures that only lead to minimums. \\ 
		
		\citet{Kholshevnikov}  & No application explicitly stated in the article, although it is obvious. & 		Theoretical procedure to transform the problem of finding the critical points of the distance between two elliptic orbits into the determination of roots of an eighth-degree trigonometric polynomial. \\  
		
		\citet{Bonnano}  & Filtering of close approaches between asteroids with Earth MOIDs smaller than \SI{0.05}{\astronomicalunit}. 		& An analytical approximation of the MOID meant mainly to obtain the   covariance matrix rather than to calculate the MOID. \\ 
		
		\citet{Gronchi}  & Evaluating the collision risk of asteroids or comets with the Solar system planets  & 		Application of geometric algebra techniques to  search for the real roots of a univariate polynomial of degree 16 that gives the critical points of the distance between two orbits using Fast Fourier Transform. \\ 
		
		\citet{Armellin}  & Close encounters between real solar system bodies & 		Solving critical values of distance between two orbits as a global optimization problem, applying a rigorous global optimizer based on Taylor models.\\ \hline
		
		\multicolumn{3}{c}{Numerical (iterative approach)}\\ \hline

		\citet{Sitarski}  & Study of parabolic comet approaches to major planets and their possible capture. & 		Numerical resolution by bisection method of the two gradient cancellation equations of the three-dimensional distance for seeking stationary points, followed by the type checking of the critical point and the selection of the global minimum. \\ 
		
		\citet{Milisavljevic}  & Assess close encounters between asteroids and  planets & 		Ingenious iterative procedure based on the orthogonality of the line of minimum distance to both orbits. \\ 
		
		\citet{Wisnioski}  & Massive calculations of MOIDs where speed prevails over precision. & 		A rapid geometrical method characterized by the fact that the certainty of hitting the MOID is inversely proportional to the calculation speed.\\ 
		
		
		\citet{Derevyanka}  & Classification of asteroids as potentially hazardous (PHA) by estimating the MOID  & 		A numerical method based on approximating the actual distance between orbits by the distance measured exclusively in each ecliptic meridian plane and using parallel programming to accelerate the computation time of the MOID. \\ 		\hline
	\end{tabular}
\end{table*}

The Space Dynamics Group (SDG) computation method of the MOID (from now on \sdg method) can be classified as numerical iterative and it is based on the following two algorithms:
\begin{description}
	\item[Algorithm 1:] For a given point $P_0$, the minimum distance $d$ to the primary ellipse $\mathfrak{e}_1$ is computed (see Fig.~\ref{fig:DPE3}).\\
	\item[Algorithm 2:] As we move $P_0$ along the secondary ellipse  $\mathfrak{e}_2$, by using Algorithm 1 we obtain a set $S$ of minimum distances to  $\mathfrak{e}_1$. The absolute minimum of set $S$ is then selected, which is the MOID  (see Fig.~\ref{fig:TCE}).
\end{description}

Both algorithms have strong theoretical foundations, as shown below.

A preliminary version of this work was presented at 2016 Stardust Final Conference on Asteroids and Space Debris \citep{Stardust}.

The remainder of this article is organized as follows: section~\ref{S:MFSDG} gives a more detailed description of both algorithms of the \sdg method, starting with its mathematical foundation. Section~\ref{S:DEVELOP} deals with the calculations made during the development of the software, which justify the choice of its individual components. Section~\ref{S:COMPARISON} includes numerical tests for execution time and accuracy. All results are compared with those of a well accredited method \citep{Gronchi}. Indirect comparisons with other methods are also carried out. Finally, section~\ref{SS:CONCL} presents the conclusions.

\section{Mathematical foundations of the \sdg method\label{S:MFSDG}}

\subsection{Distance between two confocal ellipses \label{ss:DEE3D}}

Let $\mathfrak{e}_1,\mathfrak{e}_2\subset\mathbb{R}^3$ (Euclidean 3-space) be two ellipses with a common focus $F$ (Fig.~\ref{fig:TCE}). The MOID is the minimum Euclidean distance between these ellipses considered as sets of points
\begin{equation}
    d(\mathfrak{e}_1,\mathfrak{e}_2) = \min_{E_{\textrm{1}}\in\mathfrak{e}_{\textrm{1}},E_{\textrm{2}}\in\mathfrak{e}_{\textrm{1}} } d(E_1,E_2)
\end{equation}

Distance computation is simpler if we take eccentric anomalies $u_1$ and $u_2$ of each ellipse as variables:
\begin{equation}
d(\mathfrak{e}_1,\mathfrak{e}_2) = \min_{u_1,u_2\in[0,2\pi]} d(E_1(u_1),E_2(u_2))
\end{equation}
This requires transferring vectors between four reference frames:

\noindent$\bullet$ $Fx_0y_0z_0$, the common inertial reference frame, with origin in focus $F$ and orthogonal directions given by right-handed  vectrix $\left[ \vect{i}_0\ \vect{j}_0\ \vect{k}_0 \right]$. Classical element sets $a_j, e_j, \Omega_j, i_j, \omega_j$ of both orbits refer to this frame (not shown in figures).

\noindent$\bullet$ $Fp_jq_jh_j$,  perifocal reference of orbit $j$, with directions given by the right-handed vectrix $\left[ \vect{u}_{p_j} \vect{u}_{q_j} \vect{u}_{h_j}\right]$: $\vect{u}_{p_j}$ towards pericentre and $\vect{u}_{h_j}$ along angular momentum vector. That of orbit 1 is represented in Fig.~\ref{fig:TCE}.

\begin{figure}
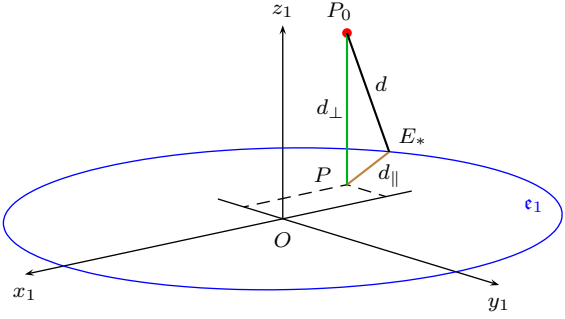

	\centering
	\vspace{5mm}
	\psset{unit=0.05\textwidth}%
	\psset{Alpha=40,Beta=15,RotSequence=zxy}%
	\pspicture(-4,-1)(5,4)%
	\small%
	\rput[tl](0.5,1){
		\pstThreeDPut(0,0,0){\rput[t](0,-0.2){$O$}}%
		\pstThreeDLine[linewidth=.5pt,linecolor=black,arrows=->](-2.5,0,0)(5,0,0)%
		\pstThreeDPut(5,0,0){\rput[t](0,-.2){$x_1$}}%
		\pstThreeDLine[linewidth=.5pt,linecolor=black,arrows=->](0,-1.5,0)(0,5,0)%
		\pstThreeDPut(0,5,0){\rput[t](0,-.2){$y_1$}}%
		\pstThreeDLine[linewidth=.5pt,linecolor=black,arrows=->](0,0,0)(0,0,3)%
		\pstThreeDPut(0,0,3){\rput[t](0,.3){$z_1$}}%

		
		\pstThreeDEllipse[linewidth=0.5pt,linecolor=blue](0,0,0)(4.25,0,0)(0,4,0)%
		
		\pstThreeDPut(1.436,0,0){%
		}%
		\pstThreeDDot[drawCoor=true,linecolor=red](-2,-0.9,2.35)%
		\pstThreeDPut(-3,-1.8,2.4){\rput[b](-0.3,-0.2){\textcolor{black}{$P_0$}}}%
		\pstThreeDPut(-3,-1.8,0){\rput[b](-0.55,-0.3){\textcolor{black}{$P$}}}%
		\pstThreeDPut(-3.7,-1.95,0){\rput[b](0.35,0.1){\textcolor{black}{$E_*$}}}%
		\pstThreeDLine[linecolor=brown](-2,-0.9,0.0)(-3.7,-1.95,0)%
		\pstThreeDPut(-3.03,-0.83,0){\rput[b](-0.2,-0.2){\textcolor{black}{$d_\parallel$}}}%
		\pstThreeDLine[linecolor=darkgreen](-2,-0.9,0.0)(-2,-0.9,2.35)%
		\pstThreeDPut(-2,-0.55,1.25){\rput[r](-.25,0){\textcolor{black}{$d_\perp$}}}%
		\pstThreeDLine[linecolor=black](-3.7,-1.95,0)(-2,-0.9,2.35)%
		\pstThreeDPut(-2.8,-1.2,1.25){\rput[bl](0,0){\textcolor{black}{$d$}}}%
		\pstThreeDPut(-3,2,0){\rput[bl](0,0){\textcolor{blue}{$\mathfrak{e}_1$}}}%
	} 
	\endpspicture
	\caption{Distance from a point to an ellipse in $\mathbb{R}^3$.}
	\label{fig:DPE3}
\end{figure}

\noindent$\bullet$ $Ox_1y_1z_1$, central frame of primary ellipse $\mathfrak{e}_1$ (see Figs.~\ref{fig:TCE} and~\ref{fig:DPE3}), with origin $O$ in the geometric centre of the ellipse, and directions parallel to the perifocal frame of this orbit.

Rotation matrix $\smat{Q}_{0j}$ from reference $0$ to reference $j$ is
\begin{alignat}{2}
\left[ \vect{u}_{p_j}\ \vect{u}_{q_j}\ \vect{u}_{h_j} \right] &= %
\left[ \vect{i}_0\ \vect{j}_0\ \vect{k}_0 \right] \,{\smat{Q}_{0j}}= \nonumber\\
&= \left[ \vect{i}_0\ \vect{j}_0\ \vect{k}_0 \right] \,\smat{R}_z(\Omega_j)
\smat{R}_x(i_j) \smat{R}_z(\omega_j)
\end{alignat}
where $\smat{R}_z(\alpha)$ is the rotation matrix of angle $\alpha$ about the axis $Oz$, in the classical Euler angle sequence.
The relative rotation matrix from the perifocal reference of  $\mathfrak e_2$ to that of  $\mathfrak e_1$ is
\begin{alignat}{2}
&\left[ \vect{u}_{p_2}\ \vect{u}_{q_2}\ \vect{u}_{h_2} \right] =
\left[ \vect{u}_{p_1}\ \vect{u}_{q_1}\ \vect{u}_{h_1} \right] \,
\overbrace{\smat{Q}_{01}^\intercal \smat{Q}_{02}}^{\smat{Q}_{12}} \\
&\smat{Q}_{12} = \smat{R}_3^\intercal(\omega_1)\; \smat{R}_1^\intercal(i_1)\; \smat{R}_2(\Omega_2\!-\!\Omega_1)\; \smat{R}_1(i_2)\; \smat{R}_2(\omega_2)
\end{alignat}
In the central frame of $\mathfrak e_1$, the position vector of point $E_2\in\mathfrak e_2$ is $\vect{OE}_2=\vect{FE}_2+\vect{OF}$, and its components are
\begin{equation}
\begin{bmatrix}  x_1^{E_2}(u_2) \\ y_1^{E_2}(u_2) \\ z_1^{E_2}(u_2)  \end{bmatrix} =
\smat{Q}_{12}\,
\begin{bmatrix}
a_2 \left(\cos u_2-e_2\right) \\%
b_2 \sin u_2 \\ %
0
\end{bmatrix}+\begin{bmatrix} a_1 e_1\\0 \\ 0 \end{bmatrix}
\end{equation}
The Euclidean distance between points $E_1$ and $E_2$ is
\begin{alignat}{2}
d(u_1, u_2) &= |\vect{OE_1}(u_1) - \vect{OE_2}(u_2)|= \nonumber\\
&=\sqrt{\sum_{{i}= 1}^3 \left(x_{{i}}^{E_1}(u_1)-x_{{i}}^{E_2}(u_2)\right)^2} \label{eq:du1u2}
\end{alignat}

Therefore, MOID computation is an unconstrained minimization of the function of  two variables given by \eqref{eq:du1u2}. The geometric meaning of the eccentric anomaly makes the function $d(u_1,u_2)$ $2\pi$-periodic in each of its two arguments, so the final domain will be restricted to $(u_1,u_2)\in[0,2\pi]\times[0,2\pi]$ (a topological torus).

\subsection{Distance between a point and an ellipse\label{ss:DPE3D}}

Let $P_0\in\mathbb{R}^3$ be a point, $\mathfrak{e}\subset\mathbb{R}^3$ an ellipse, and $u\in[0,2\pi]$ its eccentric anomaly. Distance is defined as
\begin{equation}
    d(P_0,\mathfrak{e}) = \min_{u \in [0,2\pi]} d(P_0, E(u)) = \min_{u \in [0,2\pi]} d(u)
\end{equation}

Let $P$ be the orthogonal projection of $P_0$ onto the plane of $\mathfrak{e}$ (see Fig.~\ref{fig:DPE3}). $\vect{P_0E_*}$ can be divided into components normal to the orbit plane and parallel to it
\begin{equation}
d(P_0,\mathfrak{e})^2=d_{\perp}^2 + d_{\parallel}^2=|\vect{P_0P}|^2+|\vect{PE_*}|^2
\end{equation}

where $E_*$ is the point of $\mathfrak{e}$ which gives minimum distance. Note that the plane $P_0PE_*$ is normal to $\mathfrak{e}$ at $E_*$; this property was exploited by \citet{Milisavljevic} to compute MOID.

Keeping $P_0$ fixed, the out-of-plane distance $d_{\perp}$ is constant. Therefore only the in-plane distance has to be minimized. This is the core of Algorithm~1 which provides the point $E_*$ where in-plane distance reaches its minimum value $d_{\parallel}$.

Since $Ox_1y_1z_1$ are symmetry axes of the ellipse, distance to point $P_0$ is invariant to symmetries respect to coordinate planes. We only consider the first octant of $Ox_1y_1z_1$. The same applies to the projection $P$: we only need to consider the first quadrant.

\subsection{Distance between an ellipse and a coplanar point: in-plane distance \label{ss:DPE2D}}

Let $E$ be a point of ellipse $\mathfrak e$, and $P$ a point contained in its plane. The minimum distance is
\begin{equation}
    d(P,\mathfrak{e}) = \min_{E\in\mathfrak{e}} d(P, E)
\end{equation}

and it can be computed in Cartesian coordinates or with parametric equations. The first approach gives useful insight on the number and distribution of distance minima, while the second is better suited for numerical minimization.

\subsubsection{First approach: Cartesian coordinates\label{ss:CARTESIAN}}

\begin{figure}
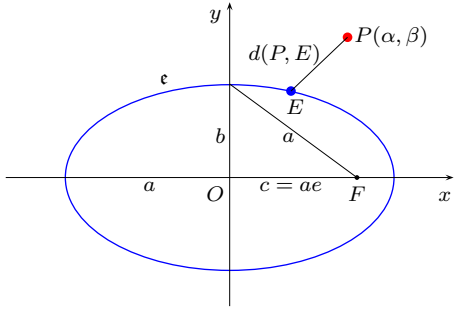

\centering
\psset{unit=0.035\textwidth}%
\pspicture(0,0)(10,8)%
\rput(5,4){%
    \psellipse[linecolor=blue,linewidth=.5pt](0,0)(3.5,2)
    \rput[br](-1.3,2){$\mathfrak{e}$}

    \rput[B](-0.3,-0.5){$O$}
    \psline[linewidth=0.35pt]{->}(-4.75,0)(4.75,0)
    \rput[Br](4.7,-.5){$x$}
    \psline[linewidth=0.35pt]{->}(0,-2.75)(0,3.75)
    \rput[B](-0.3,3.4){$y$}

    \psdot[linecolor=red](2.5,3)
    \rput(2.25,3){\rput[bl](0.4,-0.2){$P(\alpha,\beta)$}}
    \psdot[linecolor=blue](1.3,1.85)
    \rput(0.9,1.45){\rput[bl](0.3,-0.10){$E$}}
    \psline[linecolor=black,linewidth=0.35pt]{-}(1.3,1.85)(2.5,3) 
    \rput(0,2.6){\rput[bl](0.4,-0.20){$d(P,E)$}}

    \rput(-1.7,-0.2){$a$}
    \rput(-0.2,0.9){$b$}
    \rput(1.3,-0.2){$c=ae$}
    \psdot[dotscale=0.5,linecolor=black](2.7,0)
    \rput[B](2.7,-0.5){$F$}
    \psline[linecolor=black,linewidth=0.35pt]{-}(2.7,0)(0,2) 
    \rput(1.25,0.85){$a$}
}%
\endpspicture%
\caption{In-plane distances between an ellipse and a coplanar point.}
\label{fig:PRI}
\end{figure}

Let $Oxy$ be the symmetry axes of ellipse $\mathfrak{e}$ (see Fig.~\ref{fig:PRI}),  with semiaxes $a$ and $b$, parallel to $Ox$ and $Oy$, respectively. Let $(\alpha,\beta)$ be the Cartesian coordinates of point $P$ in $Oxy$. The minimization problem to solve is
\begin{gather}
d^2(P,\mathfrak{e}) =  \min_{(x,y)\in\mathfrak e} \left[(x-\alpha)^2 + (y-\beta)^2\right]\label{eq:DPEC}\\
\text{subject to } \;\dfrac{x^2}{a^2} + \dfrac{y^2}{b^2} = 1 \label{eq:RIE}
\end{gather}

We use the method of Lagrangian multipliers. Eliminating the multiplier from the objective function derivatives, we obtain
\begin{equation}
xy(a^2 - b^2) -\alpha\,a^2\,y + \beta\,b^2\, x = 0 \label{eq:RIH}
\end{equation}
which is the implicit Cartesian representation of an equilateral hyperbola $\mathfrak{h}$ of asymptotes parallel to the axes,  passing through $P$ and $O$. This property is known since the third century BC \citep{Apollonius}.

\begin{figure}
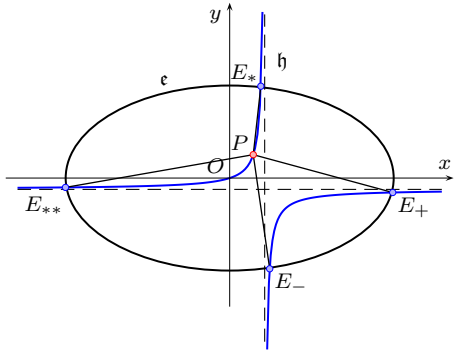

	\centering
	\psset{unit=0.035\textwidth}%
	\pspicture(0,0)(10,8)%
	\rput(5,4){%
		\psellipse[linecolor=black,linewidth=.75pt,dimen=center](0,0)(3.5,2)
		\rput[br](-1.3,2){$\mathfrak{e}$}
		\psplot[linecolor=blue,linewidth=.75pt,plotpoints=100]{-4.5}{.695}{
		  x 2. mul neg x 8.25 mul 6.125 neg add div %
		}
		\psplot[linecolor=blue,linewidth=.75pt,plotpoints=100]{.795}{4.5}{
			x 2. mul neg x 8.25 mul 6.125 neg add div %
		}
		\rput[l](1,2.5){$\mathfrak{h}$}
		\psline[linewidth=.4pt,linestyle=dashed](.7424,-3.5)(.7424,3.5)%
		\psline[linewidth=.4pt,linestyle=dashed](-4.5,-.2424)(4.5,-.2424)%
		\rput[br](-0.1,0.1){$O$}
		\psline[linewidth=0.35pt]{->}(-4.75,0)(4.75,0)
		\rput[br](4.7,.2){$x$}
		\psline[linewidth=0.35pt]{->}(0,-2.75)(0,3.75)
		\rput[B](-0.3,3.4){$y$}
		\psline[linewidth=.5pt,linecolor=black](0.5,0.5)(.8484,-1.94)%
		\psline[linewidth=.5pt,linecolor=black](0.5,0.5)(3.458,-.3087)%
		\psline[linewidth=.5pt,linecolor=black](0.5,0.5)(.6608,1.964)%
		\psline[linewidth=.5pt,linecolor=black](0.5,0.5)(-3.4825,-.1998)%
		\pscircle[linecolor=red,linewidth=0.1pt,fillstyle=solid,fillcolor=red!30](0.5,0.5){.07}
		\rput[br](0.4,0.6){$P$}
		\pscircle[linecolor=blue,linewidth=0.1pt,fillstyle=solid,fillcolor=blue!30](.8484,-1.94){.07}%
		\pscircle[linecolor=blue,linewidth=0.1pt,fillstyle=solid,fillcolor=blue!30](3.458,-.3087){.07}%
		\pscircle[linecolor=blue,linewidth=0.1pt,fillstyle=solid,fillcolor=blue!30](.6608,1.964){.07}%
		\pscircle[linecolor=blue,linewidth=0.1pt,fillstyle=solid,fillcolor=blue!30](-3.4825,-.1998){.07}%
		\rput(.8484,-1.94){\rput[tl](.1,-.1){$E_-$}}%
		\rput(3.458,-.3087){\rput[tl](.1,-.1){$E_+$}}%
		\rput(.6608,1.964){\rput[br](-.05,.1){$E_*$}}%
		\rput(-3.4825,-.1998){\rput[tr](-.05,-.2){$E_{**}$}}%
	}%
	\endpspicture%
	\caption{Apollonius' solution of stationary distances between an ellipse and a coplanar point $P$. Case with four critical distances: $d(P,E*)$ global minimum, $d(P,E\!*\!*)$ global maximum, $d(P,E+)$ local maximum, $d(P,E-)$ local minimum}
	\label{fig:E&H}
\end{figure}

Each solution of the system (\ref{eq:RIE},~\ref{eq:RIH})  is an intersection point of $\mathfrak{e}$ and $\mathfrak{h}$, where the distance to $P$ has a stationary value.
The intersections of two second order curves consist of 0, 2 or 4 points (counted with their multiplicities). Fig.~\ref{fig:E&H} shows an example with the relative disposition of both curves. Notice that
	\begin{itemize}
		\item $\mathfrak{e}$ has its centre in the origin $O(0,0)$
		\item $\mathfrak{h}$ is an equilateral hyperbola. The asymptote parallel to $Oy$ has the equation $x = \frac{\alpha}{e^2}> \alpha \ge0$; the asymptote parallel to $Ox$ has the equation $y=-\beta\frac{1-e^2}{e^2}\le 0$.
	\end{itemize}
where $e$ is the eccentricity of the ellipse.

The {upper} hyperbolic branch comes down from infinity through $P(\alpha,\beta)$,  through the origin $O$, and again to infinity towards the left. Thus $\mathfrak{h}$ and $\mathfrak{e}$ cut each other at least in two points, and system (\ref{eq:RIE},~\ref{eq:RIH}) always has at least two real roots: the problem of extreme distances always has solution \citep{Armellin}. Besides, conics are simple curves, without inflection points, so this branch can only intersect the ellipse at two points.

The {upper} hyperbolic branch is located to the left of the vertical asymptote and above the horizontal one; the same happen with points $P$ and $O$ that belong to this branch. As a consequence,
\begin{enumerate}
	\item in the limit $x \to (\frac{\alpha}{e^2})^{-}$ the {upper} branch of $\mathfrak{h}$ approaches the vertical asymptote. By continuity, it cuts the ellipse $\mathfrak{e}$ in a single point $E_*$ of the first quadrant. This is \textsl{the global minimum} of the distance.
	\item in the limit $ x \to - \infty$  the {upper} branch of $\mathfrak{h}$ approaches the horizontal asymptote. By continuity, it cuts the ellipse $\mathfrak{e}$ in a single point $E_{**}$ of the third quadrant. This is \textsl{the global maximum} of the distance.
\end{enumerate}

\begin{figure}
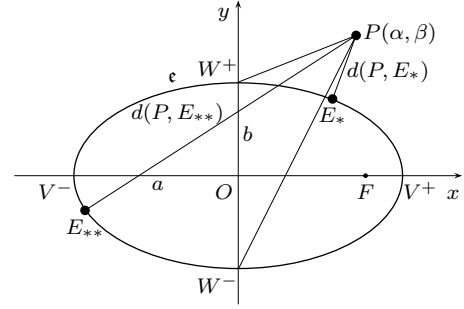

	\centering
	\psset{unit=0.035\textwidth}%
	\pspicture(0,0)(10,7)%
	\rput(5,3){%
		\psellipse[linecolor=black,linewidth=.5pt](0,0)(3.5,2)
		\rput[br](-1.3,2){$\mathfrak{e}$}
		
		\rput[B](-0.3,-0.5){$O$}
		\psline[linewidth=0.35pt]{->}(-4.75,0)(4.75,0)
		\rput[Br](4.7,-.5){$x$}
		\psline[linewidth=0.35pt]{->}(0,-2.75)(0,3.75)
		\rput[B](-0.3,3.4){$y$}
		
		\psdot[linecolor=black](2.5,3)
		\rput(2.25,3){\rput[bl](0.4,-0.2){$P(\alpha,\beta)$}}
		\psdot[linecolor=black](2,1.641)
		\rput[b](2.0,1.1){$E_*$}
		\psline[linecolor=black,linewidth=0.35pt]{-}(2.0,1.641)(2.5,3) 
		\rput[bl](2.3,2.0){$d(P,E_*)$}
		\psdot[linecolor=black](-3.25,-0.742)
		\rput[t](-3.25,-0.942){$E_{**}$}
		\psline[linecolor=black,linewidth=0.35pt]{-}(-3.25,-0.742)(2.5,3) 
		\rput[br](-0.3,1.1){$d(P,E_{**})$}
		
		\rput(-1.7,-0.2){$a$}
		\rput(0.2,0.9){$b$}
		\psdot[dotscale=0.5,linecolor=black](2.7,0)
		\rput[B](2.7,-0.5){$F$}
		\rput[Bl](3.5,-.5){$V^+$}
		\rput[Br](-3.5,-.5){$V^-$}
		\rput[tr](0,2.5){$W^+$}
		\psline[linecolor=black,linewidth=0.35pt]{-}(2.5,3)(0,2) 
		\rput[br](0,-2.5){$W^-$}
		\psline[linecolor=black,linewidth=0.35pt]{-}(2.5,3)(0,-2) 
	}%
	\endpspicture%
	\caption{Extreme distances vs. those to vertices $W^+$ and $W^-$.}
	\label{fig:DISTANCES}
\end{figure}

Let $V^+$, $V^-$, $W^+$ and $W^-$ be the vertices  of $\mathfrak{e}$ (see Fig.~\ref{fig:DISTANCES}). The following geometric relations are useful to identify the global extreme distances
\begin{alignat}{2}
d(P,W^+) &=  \sqrt{\alpha^2+(b-\beta)^2} \\
d(P,W^-) &=  \sqrt{\alpha^2+(b+\beta)^2} \\
d(P,E_*) &< d(P,W^+) < d(P,W^-) < d(P,E_{**}) \label{ine:DEDC}
\end{alignat}

\begin{figure}
	\centering
	\includegraphics[width=60mm, clip]{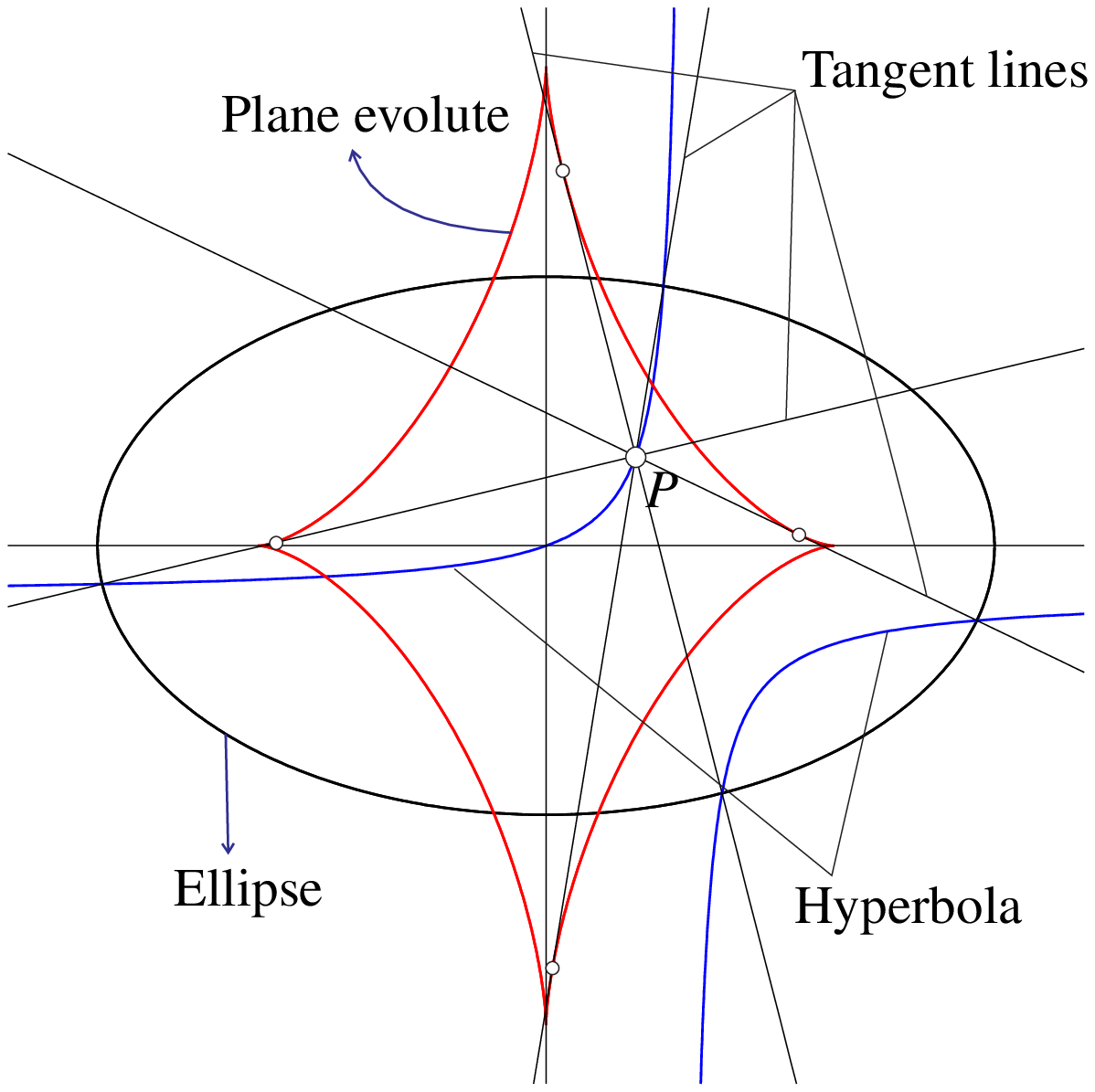} 
	\caption{Tangent lines to the plane evolute of an ellipse from an interior point, $\Pi(\alpha,\beta)<0$.}
	\label{fig:PIEP}
	
	\includegraphics[width=60mm, clip]{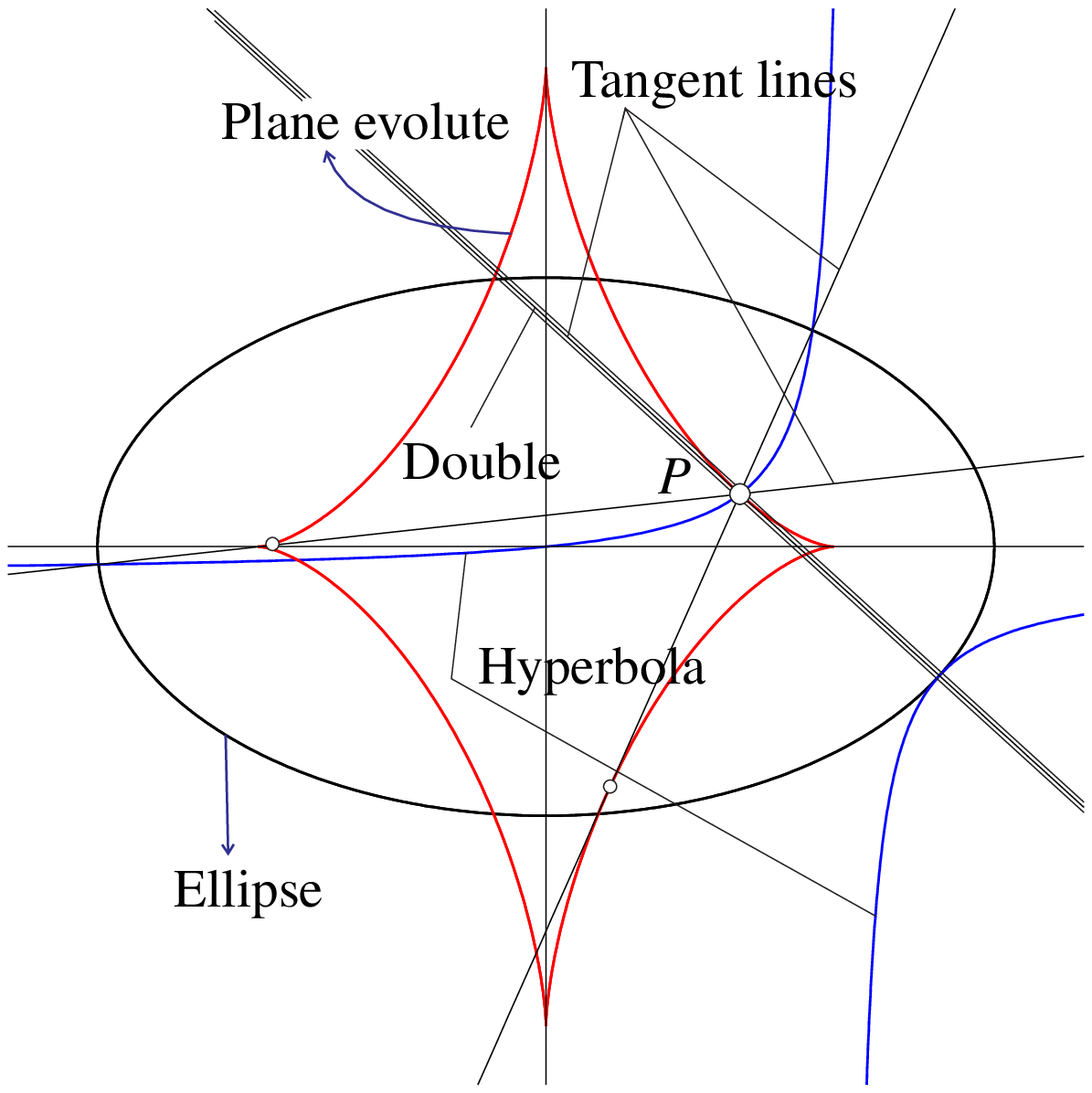} 
	\caption{Tangent lines to the plane evolute of an ellipse from a point on the evolute itself,  $\Pi(\alpha,\beta)=0$.}
	\label{fig:PSEP}
	
	\includegraphics[width=60mm, clip]{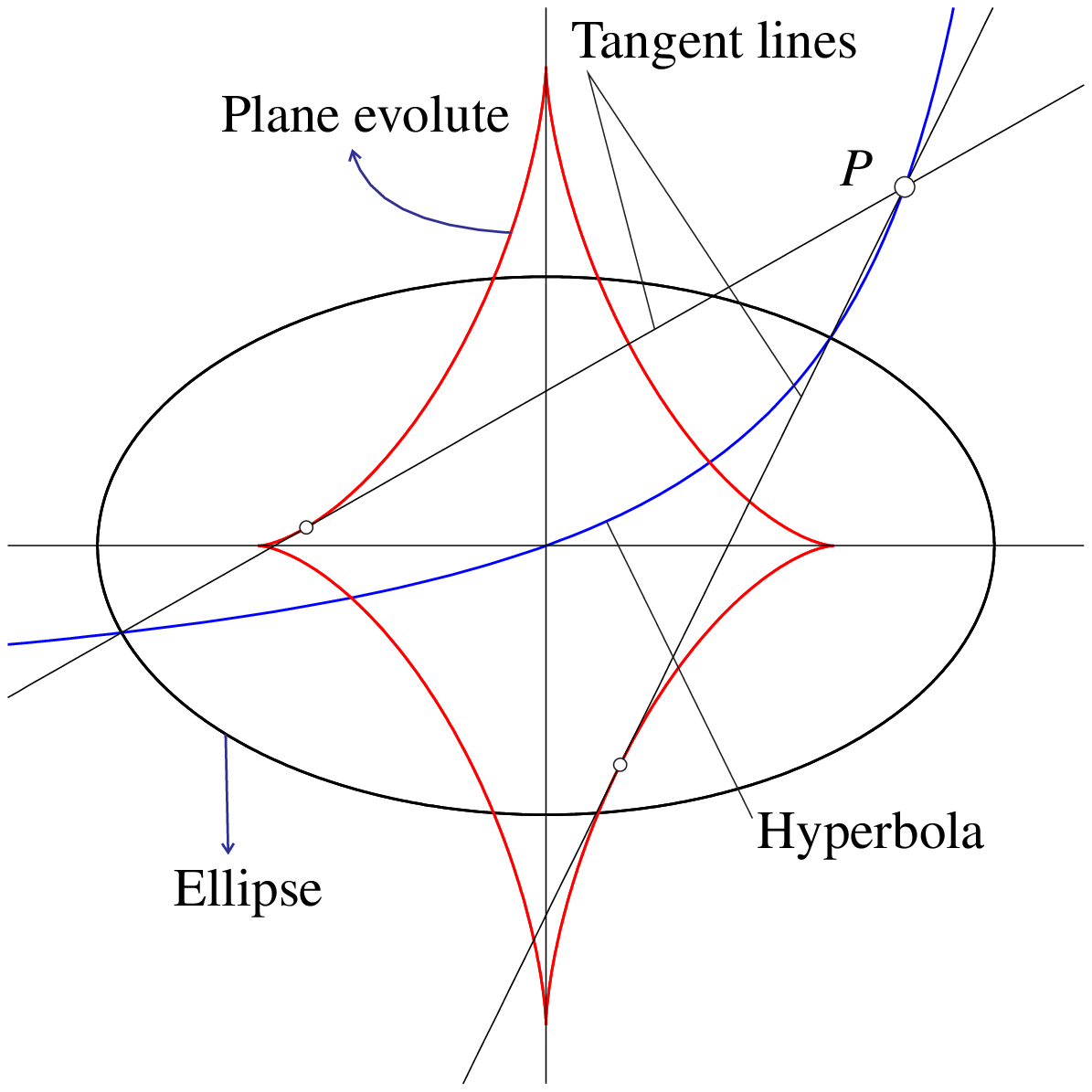} 
	\caption{Tangent lines to the plane evolute of an ellipse from an exterior point,  $\Pi(\alpha,\beta)>0$.}
	\label{fig:PEEP}
\end{figure}

The {lower} hyperbolic branch is located to the right of the vertical asymptote, and below the horizontal one. It can only cut the ellipse in the fourth quadrant: two cuts, one (tangent), or none.  Fig.~\ref{fig:E&H} shows a case with two cuts, $E_-$ and $E_+$. The first one gives a relative minimum and the other  a relative maximum, unless both join into a single tangent point. Figs.~\ref{fig:PIEP}--\ref{fig:PEEP} show the three possibilities. In any case, they satisfy
\begin{alignat}{2}
d(P,E_*) &< \min\left(d(P,V^+), d(P,W^-)\right) < d(P,E_-) \le\nonumber\\
&\le  d(P,E_+) < d(P,E_{**}) \label{ine:DEDV}
\end{alignat}

The intersection points of $\mathfrak e$ and $\mathfrak h$ can be obtained by eliminating $y$ between Eqs. \eqref{eq:RIH} and \eqref{eq:RIE}:
\begin{multline}
x^4\,e^4 -2\,\alpha\,e^2\, x^3 - (a^2\,e^4 + \beta^2\,e^2 - [\alpha^2 + \beta^2])\,x^2 +\\
+ 2\,\alpha\,e^2\, a^2\,x -a^2\,\alpha^2=0 \label{eq:QE}
\end{multline}

Eq. \eqref{eq:QE} is a quartic with real coefficients, which in general can have 0, 2, or 4 real roots (counting the multiplicity index). As shown by the intersections with the hyperbola, in this case there are at least 2 real roots. The exact number can be determined by geometric considerations.

It is well known that, for critical distances, $\vect{PE}$ is normal to the ellipse $\mathfrak e$ and tangent to its plane evolute $\mathfrak p$ -- see~\citet{Milisavljevic}. Since all normals are tangent to the evolute, there are as many real roots of \eqref{eq:QE} as tangents to $\mathfrak{p}$ from $P$. The number depends on the position of $P$ relative to $\mathfrak{p}$.
	
	The implicit equation of $\mathfrak p$ (see Figs.~\ref{fig:PIEP}-\ref{fig:PRP}) is	
	\begin{equation}
	\Pi(x,y) =  x^{\frac{2}{3}} + \left (y\,\sqrt{1-e^2} \right )^{\frac{2}{3}} - a^{\frac{2}{3}}\,e^{\frac{4}{3}} = 0
	\label{eq:RIEP}
	\end{equation}
	Substituting the coordinates   $(\alpha,\beta)$ of $P$ into  $\Pi$ shows its relative position.
\begin{enumerate}
	\item If $P$ is {inside} $\mathfrak{p}$ (Fig.~\ref{fig:PIEP}), $\Pi(\alpha, \beta) <0$ and there are four different tangents to $\mathfrak{p}$, unless $P$ is upon one of the axes. In this case there are only three, but one is the axis itself, which is tangent at two points. Therefore, there are four roots of Eq. \eqref{eq:QE}.
	
	\item If $P$ is {on} $\mathfrak{p}$ (Fig.~\ref{fig:PSEP}), $\Pi(\alpha, \beta)  =0$ and there are only three different tangents from $P$. So the number of different real roots of Eq.~\eqref{eq:QE} is three but one, the tangent at $P$ itself, is double (two tangents to the same branch join into one).
	
	\item If $P$ is {outside} $\mathfrak{p}$ (Fig.~\ref{fig:PEEP}),  $\Pi(\alpha, \beta)  >0$ and only two tangents to $\mathfrak{p}$ can be traced from $P$; then the number of different real roots of Eq.~\eqref{eq:QE} is two.
\end{enumerate}
Two consequences can be drawn from the previous results:
(a) the problem always has at least one solution with minimum distance and one solution with maximum distance; and (b) there may be two different real roots that give minimum distance because of the symmetries.

\subsubsection{Second approach: parametric representation}

\begin{figure}
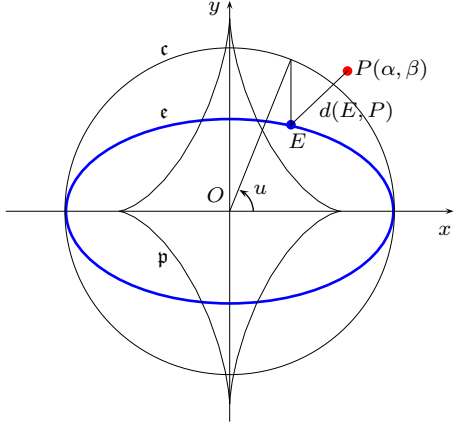

\centering
\psset{unit=0.035\textwidth}%
\pspicture(0,0)(10,9)%
\rput(5,4.5){%
    \psellipse[linecolor=blue,linewidth=1pt](0,0)(3.5,2)
    \rput[br](-1.3,2){$\mathfrak{e}$}

    \rput[br](-0.1,0.2){$O$}
    \psline[linewidth=0.35pt]{->}(-4.75,0)(4.75,0)
    \rput[br](4.7,-.5){$x$}
    \psline[linewidth=0.35pt]{->}(0,-4.5)(0,4.5)
    \rput[br](-.2,4.2){$y$}

    \psdot[linecolor=red](2.5,3)
    \rput(2.25,3){\rput[bl](0.4,-0.2){$P(\alpha,\beta)$}}


    \psdot[linecolor=blue](1.3,1.85)
    \rput(0.9,1.45){\rput[bl](0.35,-0.10){$E$}}
    \psline[linewidth=0.35pt](1.3,1.8)(1.3,3.25)
    \psline[linewidth=0.35pt](0,0)(1.3,3.25)
    \psline[linewidth=0.35pt](1.3,1.85)(2.5,3) 
    \rput(1.9,2.1){\rput[bl](0,-0.2){$d(E,P)$}}

    \psarc[linewidth=.5pt]{->}(0,0){0.5}{0}{66} %
    \put(0,0){\rput[t](0.65,0.55){$u$}}

    \parametricplot[linewidth=.35pt,linecolor=black]{0}{360}{%
    t cos dup dup mul mul 2.357 mul t sin dup dup mul mul 4.125 mul neg %
    }%
    \rput[br](-1.3,-1.2){$\mathfrak{p}$} 




    \pscircle[linewidth=.2pt](0,0){3.5}
    \rput[r](-1.3,3.45){$\mathfrak{c}$}
}%
\endpspicture%
\caption{Geometry of the in-plane distance problem in parametric equations.}
\label{fig:PRP}
\end{figure}

Consider the  parametric equations of the ellipse $\mathfrak{e}$ in terms of the eccentric anomaly $u \in \mathbb{R}$, as shown in Fig.~\ref{fig:PRP}
\begin{equation}
x(u)=a\cos(u)\;;\quad  y(u)=b\sin(u)
\end{equation}

The distance $d(P,E)$ is
\begin{equation}
d(u;a,b,\alpha,\beta)=\sqrt{ (\alpha-a\cos u)^2 +(\beta-b\sin u)^2}
    \label{eq:DPEP}  
\end{equation}
where $u$ is the independent variable and $a,b,\alpha,\beta$ are the four parameters of the problem.  For any parameter values, the function $d(u)$ is:
\begin{enumerate}
  \item $2\pi$-periodic; as a consequence $u\in[0,2\pi]$;
  \item always non-negative, zero if and only if $P\in\mathfrak{e}$.
\end{enumerate}

Using the squared distance simplifies the computation of stationary points $u^*$, since the square root is avoided, and the stationary points and their character are preserved (see appendix~\ref{APP:PROPERTIES}). Stationary points are related to zeros of the first derivative.	

We introduce the univariate function
\begin{align}
f(u;a,b,\alpha,\beta)&=\td{\left[d^2(u;a,b,\alpha,\beta)\right]}{u}=\nonumber  \\
&=\alpha a \sin u - \beta b \cos u - c^2 \sin u \cos u  \label{eq:FDEU}
\end{align}	
whose real roots provide the eccentric anomalies of the points of ellipse $\mathfrak{e}$  with stationary distance to  $P(\alpha,\beta)$. These are the same points obtained from Eq. \eqref{eq:QE}, but here the variable is the parameter $u$.

\subsubsection{Study of minimum distance cases}

Let $u^*$ be any real root of Eq. \eqref{eq:FDEU}, that is, an extremum of distance. Let $u_*$ be a real root with global minimum distance (there could be more than one due to symmetries). From now on we will assume that $P$ is in the first quadrant of $Oxy$, that is, $\alpha\ge 0$, $\beta\ge0$. We can consider two types of solutions:
\begin{enumerate}
	\item Immediate closed-form solutions, such as a circular orbit, or $P$ upon one of the principal axes. These are studied in appendix~\ref{APP:SOLUTIONS}, and summarized in Table~\ref{tab:CFS}.
	\item Those requiring the application of classic quartic polynomial solution, or numerical methods.
\end{enumerate}

\begin{table*}
  \centering
  \caption{Cases of minimum distance between an ellipse and a coplanar point with closed-form solution.}
  \label{tab:CFS}
\begin{tabular}{cccccc}
  \hline
  \phantom{\Big(}Case & Subcase & \multicolumn{1}{c}{$x^{E_*}$} & \multicolumn{1}{c}{$y^{E_*}$} &
  \multicolumn{1}{c}{$u_*$} & \multicolumn{1}{c}{$\Delta(\alpha,\beta)$ \eqref{eq:DPEP}}\\
  \hline
  $a=b=R$ & & $\frac{R\alpha}{\sqrt{\alpha^2+\beta^2}}$ & $\frac{R\beta}{\sqrt{\alpha^2+\beta^2}}$ &
  $\arctan\left(\frac{\beta}{\alpha}\right)$ & $|\sqrt{\alpha^2+\beta^2}-R|$ \\ \hline
  \multirow{4}{1cm}{$a>b$}
  & \phantom{\Big(} $\alpha=0, \beta=0$   & $0$ & $\pm b$ & $\pm \frac\pi 2$ &  $b$ \\
  & \phantom{\Big(} $\alpha=0, \beta>0$ & $0$ & $b$ & $\frac\pi 2$ & $|\beta-b|$\\
  & \phantom{\Big(} $\alpha>0, \beta=0 \left\{
  \begin{array}{l}
  \text{if } \alpha> e^2 a\\
  \text{if } \alpha\le e^2 a
  \end{array}\right.$
  & $\begin{array}{c} a\\ \frac\alpha{e^2}   \end{array}$
  & $\begin{array}{c} 0\\ a\sqrt{1-e^2}\sqrt{1-\frac{\alpha^2}{e^2a^2}}  \end{array}$
  & $\begin{array}{c} 0\\ \arccos(\frac{\alpha}{ae^2})   \end{array}$
  & $\begin{array}{c}
  |\alpha-a|\\
  \sqrt{(1-e^2)(a^2-\frac{\alpha^2}{e^2})}
  \end{array}$
  \\
  \hline
\end{tabular}
\end{table*}

We will focus on the second type, with  $\alpha>0, \beta>0, e>0$. Each method will be studied separately.

\subsubsection{Classic quartic polynomial solution\label{ss:algebr}}

If relations
\begin{equation}
\cos u = \dfrac{1}{\sqrt{1+t^2}}, \quad \sin u = \dfrac{t}{\sqrt{1+t^2}}, \quad t = \tan u
\end{equation}
are introduced in \eqref{eq:FDEU} and squared
\begin{equation}
(a\, \alpha\, t - b\, \beta)^2 - \dfrac{a^4 e^4 t^2}{1 + t^2} =0
\end{equation}
we obtain the quartic
\begin{equation}\label{eq:XX}
  a^2 \alpha^2\, t^4 - 2 \,a\, b \,\alpha \beta\,t^3 + \varpi\, t^2  -2\,a\,b\,\alpha \beta\, t + b^2 \beta^2 = 0
\end{equation}
with a symmetric structure, and where $\varpi$ is given by
\[ \varpi = a^2\alpha^2 +b^2 \beta^2 - a^4 e^4 \]
Solving this quartic equation accurately is numerically difficult for small values of $\alpha$. In that case, the change $t=1/z$ leads to a similar quartic in the variable $z$
\begin{equation}\label{eq:YY}
  a^2 \alpha^2 - 2 \,a\, b \,\alpha \beta\,z +  \varpi\, z^2 -2\,a\,b\,\alpha \beta\, z^3 + b^2 \beta^2 z^4 = 0
\end{equation}

If relations
\begin{equation}
\cos u = \dfrac{1 -m^2}{1+m^2}, \quad \sin u = \dfrac{2\,m}{1+m^2}, \quad m = \tan \dfrac{u}{2}
\end{equation}
are introduced in \eqref{eq:FDEU} and squared, we obtain
\begin{equation}
m^4 + (\Lambda + \Sigma) m^3 +(\Lambda - \Sigma) m - 1 =0 \label{eq:Q1}
\end{equation}
where the positive parameters $(\Lambda, \Sigma)$ are given by
\begin{equation}
\Lambda = \dfrac{2\,a\alpha}{b\beta}, \qquad \Sigma = \dfrac{2\,a^2e^2}{b\, \beta} \label{eq:QPAR1}
\end{equation}

The coefficients in $m^3$ and $m$ of Eq. \eqref{eq:Q1} can become very large when $\beta$ is very small. In that case, a change of variable improves numerical accuracy. If the relations
\begin{equation}
\sin u = \dfrac{1 -q^2}{1+q^2}, \quad \cos u = \dfrac{2\,q}{1+q^2}, \quad q = \tan \left (\dfrac{\pi}{4} -\dfrac{u}{2} \right )
\end{equation}
are introduced into \eqref{eq:FDEU} and squared, the following equation is obtained
\begin{equation}
q^4 + (\tilde{\Lambda} - \tilde{\Sigma}) q^3 +(\tilde{\Lambda} + \tilde{\Sigma}) q - 1 =0 \label{eq:Q3}
\end{equation}
where the positive parameters $(\tilde{\Lambda}, \tilde{\Sigma})$ are given by
\begin{equation}
\tilde{\Lambda} = \dfrac{2\,b\beta}{a\alpha} = \dfrac{4}{\Lambda}, \qquad  \tilde{\Sigma} = \dfrac{2\,ae^2}{\alpha} = \dfrac{1}{2} \Sigma\tilde{\Lambda} \label{eq:QPAR3}
\end{equation}

The classic solution can be applied to any of the above quartic polynomials.\vspace{\topsep}

In general, the quartic equations \eqref{eq:QE}, \eqref{eq:XX}, \eqref{eq:YY}, \eqref{eq:Q1}, and \eqref{eq:Q3} have four complex solutions that can be obtained by explicit formulas, but the useful solutions are real.

The analytical solution of a quartic equation has been coded and tested, but it is slower than the iterative approach, and less accurate (see Table~\ref{tab:AvsI} in subsection~\ref{ss:AvsI} below). Therefore this approach was discarded.

\subsubsection{Numerical iterative method\label{ss:numeriter}}

When $P(\alpha,\beta)$ is in the first quadrant and outside of the coordinate axes, Eq.~\eqref{eq:FDEU} gives
\begin{equation*}
  f(0) = - \beta\, b < 0 \,; \quad %
  f\left(\tfrac{\pi}{2}\right) = \alpha\,a > 0
\end{equation*}
Since $f(u)$ is a continuous function in the closed interval $[0,\frac{\pi}{2}]$, Bolzano's theorem assures that there is, at least, one root $ u_* \in [0,\frac{\pi}{2}] $.

As shown in section~\ref{ss:CARTESIAN}, there is only one root $u_*$ in the open interval $(0,\frac{\pi}{2})$. It is a simple root because $f'(u_*)\not=0$ and it defines the ellipse point with minimum distance to $P$, so $f'(u_*)>0$.

Therefore, the problem of computing the minimum distance between an ellipse $\mathfrak{e}$ and a coplanar point $P(\alpha>0, \beta>0)$ is reduced to finding the root of the univariate function $f(u)$ in the open interval $(0, \frac{\pi}{2})$, because the other cases ($\alpha=0$ and/or $\beta = 0$) have closed-form trivial solutions.

The function $f(u)$ from Eq.~\eqref{eq:FDEU}, rewritten as
\begin{equation} \label{eq:fu}
f(u)=\alpha a \sin u - \beta b \cos u - \dfrac{c^2}{2}  \sin 2 u
\end{equation}
is non-linear and smooth (of class $\textrm{C}^{\infty}$ in its definition domain). Its derivatives can be easily computed, so high-order iterative methods can be used.

Being transcendental, an iterative root-finding method must be used. Convergence is assured: Brouwer's fixed-point theorem states that, for any continuous function $\displaystyle f$ mapping a compact convex set to itself, there is a point $\displaystyle x_{0}$ such that $\displaystyle f(x_{0})=x_{0}$ -- see \citet{agarwal2001fixed}. To apply the fixed-point theorem, we rewrite it as
\begin{alignat}{2}
u &= \arctan{\left( \frac{\beta\sqrt{1-e^2}+a e^2\sin u}{\alpha}\right)} = \Phi(u) \label{eq:PHI}
\end{alignat}
$\Phi(u)$ satisfies the two conditions of Brouwer's theorem:
\begin{enumerate}
	\item $\Phi(u)$ is obviously continuous: $\Phi(u)\in\mathrm{C}^{\infty}$
	\item $\Phi([0,\frac\pi 2]) \subseteq [0,\frac\pi 2]$.
\end{enumerate}
To prove the second statement, we see first the trivial cases  $(ae\alpha=0)$, where $\Phi$ is constant and the fixed point $u_o$ is:
\begin{itemize}
    \item $\text{If } \alpha=0\ (P \in Oy)       : \Phi(u)=\frac\pi 2\ \text{(const.)} \implies u_0=\frac\pi 2$ \\
	\item $\text{If } e=0\ (\mathfrak{e}\text{ circle}) : \Phi(u)=\arctan(\frac{\beta}{\alpha})\ \text{(const.)} \implies$ \\
	$\phantom{m}$ \hfill $\implies u_0 =\arctan(\frac\beta\alpha)$\\
	\item $\text{If } a=0\ (\mathfrak{e} \equiv O) : \Phi(u)=\arctan{(\frac\beta\alpha)}\ \text{(const.)} \implies $ \\
	$\phantom{m}$ \hfill $\implies u_0 =\arctan{(\frac\beta\alpha)}$\\
\end{itemize}
For non trivial cases  $\forall \alpha>0,\beta\ge 0, a>0, e>0$, we must study $\Phi$ and its derivative:
\begin{alignat}{2}
0 &\le \Phi(0) = \arctan{\left(\frac{\beta\sqrt{1-e^2}}{\alpha}\right)} \le \nonumber\\
&\le  \arctan{\left(\frac{\beta\sqrt{1-e^2}+a e^2}{\alpha}\right)} = \Phi\left(\frac\pi 2\right) \le \frac\pi 2 \label{eq:bounds}\\
0 &< \displaystyle\Phi'(u) = \displaystyle\frac{\alpha a{e}^{2}\cos \left( u \right) }
{ \alpha^2 + \left( \beta\,\sqrt {1-{e}^{2}}+a{e}^{2}\sin \left( u \right) \right)^{2} } \;\; \forall u \in\left[0,\tfrac{\pi}{2}\right)\label{eq:monotonic}
\end{alignat}
Only in the upper limit $u=\frac \pi 2$ is $\Phi^\prime=0$, so we can conclude
\begin{enumerate}
	\item $\Phi(u)$ increases monotonically \eqref{eq:monotonic}, and\\
	\item $\Phi([0,\frac\pi 2])$ is bounded inside $[\Phi(0),\Phi(\frac\pi 2)] \subseteq [0,\frac\pi 2]$ \eqref{eq:bounds}
\end{enumerate}
Therefore, $\exists ! u_0\in [0,\frac\pi 2]$ such that $u_0 = \Phi(u_0)$.

Instead of simply iterating $ u=\Phi(u)$, a high convergence method is more efficient. Some have been considered: Newton--Raphson,  Halley's \citep{Wesstein}, and Improved Newton--Raphson \citep{Danby}. The last two are always faster. However, the specific machine determines which is the best, because of a balance between speed of convergence and number of function evaluations. A quantitative comparison will be shown in subsection~\ref{ss:HvsD}. As an example, appendix~\ref{APP:HALLEY} collects Halley's method characteristics, with its advantages and the pitfalls to avoid.

\subsection{\sdg method}

The approach used in this paper transforms the complex problem of minimization in two variables with ten parameters into a set of simpler minimization problems, each in one variable, with easier mathematical treatment. The objective is speeding up  computation.

The main idea of the method is to separate the two components of the distance between a point and an ellipse in Euclidean 3-space:   normal to the ellipse plane and contained in it. The procedure is as follows:
\begin{description}
\item[{Algorithm 1:}]{finds minimum distance of an arbitrary point $P_0$ to the primary orbit $\mathfrak e_1$.}
    \begin{enumerate}
    \item{$P_0$ is projected onto the orbital plane of $\mathfrak e_1$ as $P$}
    \item{The minimum distance from $P$ to $\mathfrak e_1$ and the corresponding eccentric anomaly are found. It uses Halley's  iterative method to find the root, which corresponds to the minimum value. }
    \item{Finally, in-plane and normal components are combined to find the distance.}
    \end{enumerate}
\item[{Algorithm 2:}] {
	MOID computation is reduced from a two-variable problem to a one-variable function minimization. Steps are:}
\end{description}
\begin{enumerate}
	\item{Discretise secondary orbit $\mathfrak e_2$, call Algorithm 1 to compute minimum distance to $\mathfrak e_1$ for each anomaly, and bracket local minima.}
	\item{Refine by a closed interval search algorithm.}
	\item{Select global minimum between the former values.}
\end{enumerate}

The detailed steps are:  define a uniform grid of $N+1$ values $\{u^0_2, u^1_2, \dots, u^N_2\}$ in the interval $u_2 \in [0,2\pi]$ where $u_2$ is the eccentric anomaly of $\mathfrak{e}_2$. The constant interval size is $ \Delta u_2 = \frac{2\pi}{N}$. A corresponding grid of $N$ image points $\{E^0_2,E^1_2, \dots, E^{N-1}_2\}$ is defined in $\mathfrak{e}_2$. The last value, $u_2^N$, is ignored because $E^0_2 \equiv E^N_2$. The choice of $N$ will be discussed in subsection~\ref{ss:gridsize}.

For each $u_2$, we compute the minimum distance to $ \mathfrak{e}_1$ through Algorithm 1 -- as described in subsection~\ref{ss:DPE3D} -- to obtain a reticular approximation of the function
\begin{equation}
\displaystyle\Phi(u_2)=\min_{u_1\in [0,2\pi]} d(u_1, u_2)
\end{equation}
which is used to bracket the abscissas of relative minima of $d(u_1,u_2)$ within an interval.

The abscissa of each local minimum is determined by decreasing the size of the uncertainty interval through golden section search. The smallest local minimum is the global one, that is, the MOID.

\section{Software development\label{S:DEVELOP}}

The NEODyS database corresponding to epoch March 23rd, 2018 = MJD 58200, from the Near Earth Objects - Dynamic Site, has been used to compute the Earth MOID.

Development includes the selection of the best algorithms for each component, and optimisation of free parameters.

\subsection{Algorithm 1: algebraic vs iterative\label{ss:AvsI}}

First, a study was made on how to calculate the distance in the plane. The approaches described in subsections~\ref{ss:algebr} and~\ref{ss:numeriter} have been compared. The complete database has been computed 20 times to minimize the effect of background machine load, and find the average unit time. Test results show:
	\begin{itemize}
		\item The algebraic method is 66 per cent slower than the iterative approach, as shown in Table~\ref{tab:AvsI}. This may be due to the arithmetic of complex numbers, frequent calls to slow functions, and many conditional bifurcations on the code.
		\item In a few cases, the characterization of a root as minimum or maximum fails at double precision. The accumulation of rounding and truncation errors leads to wrong negative radicands and other problems. Quadruple precision has been necessary to accurately cover all cases, making the algebraic method even slower.
	\end{itemize}

Consequently, the iterative method was selected.

\begin{table}
  \centering
  \caption{Unit execution times for in-plane distance algorithms: algebraic quadric solution vs Halley's iterative method. }
  \label{tab:AvsI}
  \begin{tabular}{crr}
     \hline
& \multicolumn{2}{c}{Average unit time} \\
& \multicolumn{2}{c}{\si{\micro\second}} \\
Samples & \sdg algebraic	& \sdg iterative  \\
\hline
\#1 &		55.62	& 33.73 \\
\#2 &		55.30	& 33.76  \\
\#3 &		55.36	& 33.71  \\
\#4 &		56.74	& 33.22  \\
\#5 &		55.32	& 33.10  \\
\#6 &		56.05	& 33.22  \\
\#7 &		56.49	& 33.09  \\
\#8 &		55.66	& 34.30  \\
\#9 &		55.50	& 33.72  \\
\#10&		55.01	& 33.26  \\
\hline
mean &		55.705	& 33.511 \\
st.dev. &	0.555 &	0.394 \\
\cline{1-1}
\multicolumn{1}{c}{Rel. Diff. to iterative} & 0.6623 & 0 \\
\hline
Sample size &20$\times$17648 &	20$\times$17648	 \\
     \hline
   \end{tabular}
\end{table}

\subsection{Iterative procedure: Halley vs Danby\label{ss:HvsD}}

We still have to choose the iterative method. Low order ones such as Newton--Raphson have been outright discarded. As mentioned above, Halley's \citep{Wesstein} and Danby's \citep{Danby} methods have been tested for incorporation into the \sdg code. As in the previous section, the whole database of NEOs has been computed 20 times, to find the unit average time.
	
Comparisons have been carried out in an Intel I7-4771 PC @ \SI{3.5}{\giga\hertz} with \SI{16}{\gibi\byte} of RAM. Table~\ref{tab:PC} shows mean times for each MOID in the database, and Halley's method is slightly faster. Later on, when running \sdg method against Gronchi's in a different machine, Danby's  was ahead (see Table~\ref{tab:SM} in subsection~\ref{SS:GRONCHI} below).

Results show that the most appropriate iteration method depends substantially on the technical characteristics of the computer used. Danby's method uses fewer iterations, but more function calls for each. Conversely, Halley's needs more iterations, but with fewer function calls. Thus, a higher frequency favours Halley, but differences in architecture may favour Danby. It may be necessary to test both on the target machine.

Table~\ref{tab:NEOS_ITERATION} shows the iterations needed to compute the MOID of the first 30 database objects, with seed  $\arctan(\frac{\alpha}{\beta})$, in the same machine as Table~\ref{tab:PC}.
	\begin{enumerate}
		\item 3244 Halley iterative processes are needed, with an average of 1.73 iterations per process. In most cases (72.66 per cent) two iterations are needed while in a very small part (0.31 per cent) up to three iterations are needed.
		\item 3242 Danby iterative processes are needed, with an average of 1.005 iterations per process. In most cases (99.51 per cent) one iteration is needed while in a very small part (0.49 per cent) up to two iterations are needed.
	\end{enumerate}
	
Comparing the columns of Table~\ref{tab:NEOS_ITERATION} the advantage of Danby's method over Halley's in convergence speed is clear.

\begin{table}
\centering
\caption{Results of unit execution times of the two iterative procedures tested for Algorithm 1 in the \sdg method:  Halley vs Danby.}
\label{tab:PC}
\begin{tabular}{crr}
     \hline
& \multicolumn{2}{c}{Average unit time} \\
& \multicolumn{2}{c}{\si{\micro\second}} \\
Samples & Halley	& Danby \\
\hline
\#1     & 33,73     & 35,53 \\
\#2	    & 33,76     & 35,77 \\
\#3     & 33,71     & 35,83 \\
\#4     & 33,22	    & 35,65 \\
\#5     & 33,10	    & 35,53 \\
\#6     & 33,22	    & 35,58 \\
\#7     & 33,09	    & 35,56 \\
\#8     & 34,30	    & 35,66 \\
\#9     & 33,72	    & 35,56 \\
\#10    & 33,26 	& 35,62 \\
\hline
mean	& 33,511    & 35,629 \\
st.dev.	& 0,394     & 0,102 \\
\cline{1-2}
\multicolumn{2}{c}{Rel. diff to Halley} &	0,0632 \\
\hline
Sample size	 & 20$\times$17648 &	20$\times$17648 \\
\hline
   \end{tabular}
\end{table}

\begin{table}
\centering
\caption{Number of Halley and Danby iterative processes and number of iterations for the 30 first objects of NEODyS database.}
\label{tab:NEOS_ITERATION}
\begin{tabular}{rrlrl}
\hline
            & \multicolumn{2}{c}{Halley} & \multicolumn{2}{c}{Danby} \\
No.         & Abs.  & Rel.   & Abs.  & Rel. \\
Iterations	& Freq.	& Freq.  & Freq. & Freq.\\
\hline
1           & 877	& 0.2703 & 3226	 & 0.9951 \\
2           & 2357	& 0.7266 & 16	 & 0.0049 \\
3           & 10	& 0.0031 & 0  	 & 0.0000\\
\hline
TOTAL :		& 3244	& 1.0000 & 3242	 & 1.0000\\
AVERAGE : 	& \multicolumn{2}{c}{1.733 \text{iter/proc}} & \multicolumn{2}{c}{1.005 \text{iter/proc}}\\
\end{tabular}
\end{table}

\subsection{Iteration tolerance of Algorithm~1}
One or more tolerance values must be selected to end the iterations. Two examples of stop conditions are
\begin{equation}
    |x_{{i}}-x_{{i-1}}|/|x_{{i}}| < x_{\textrm{TOL}};\quad
    |f_{{i}}-f_{{i-1}}|/|f_{{i}}| < y_{\textrm{TOL}};
\end{equation}
The values of these tolerances determine the  error admitted and must obviously be greater than the machine epsilon ($\displaystyle\epsilon$): $x_{\textrm{TOL}} > \displaystyle\epsilon, \ y_{\textrm{TOL}} > \displaystyle\epsilon$. In C++ floating point arithmetic with double precision $\displaystyle\epsilon = 2^{-52} \approx \SI{2.22E-16}{}$.

These values influence the average computation times. In \sdg method, $x_{\textrm{TOL}}=y_{\textrm{TOL}}$ has been chosen for simplicity, and some studies have been carried out to measure the influence of this value on computation times. Table~\ref{tab:TIT} collects the most significant results. Since the number of iterations carried out by \sdg method is small, the influence of these tolerances is very low as can be seen in Table~\ref{tab:TIT}.

\begin{table}
\centering
\caption{Influence of error tolerance on average computation times.}
\label{tab:TIT}
\begin{tabular}{llc}
\hline
\multicolumn{1}{c}{\bf{Tolerance}}  & \multicolumn{2}{c}{\bf{Unit Time}}\\
\multicolumn{1}{c}{Non-dim.}       & \multicolumn{2}{c}{\si{\micro\second}} \\
       & {MEAN}	& {STD. DEV.}\\
\hline
\SI{1.00E-14}{}	& 30.6761	& 0.198927374 \\
\SI{7.50E-15}{}	& 31.1439	& 0.373606284 \\
\SI{5.00E-15}{}	& 31.5604	& 0.391580615 \\
\SI{2.50E-15}{}	& 31.9417	& 0.182431753 \\
\SI{1.00E-15}{}	& 32.9995	& 0.375676057 \\
\SI{5.00E-16}{}	& 33.5142	& 0.251684459 \\
\hline
\end{tabular}
\end{table}

Finally, $x_{\textrm{TOL}}=2\epsilon \approx$ \SI{4.44E-16} has been adopted, more restrictive than the value of \SI{1E-14} chosen by other authors.

\subsection{Grid size of Algorithm 2\label{ss:gridsize}}

It is well known -- see \citet{Gronchi2002}, \citet{Milisavljevic}, or \citet{Wisnioski} -- that there can be up to 4 minimum values of the distance between the points of two confocal ellipses. For the method to be fast and accurate, grid size $N$ has to be the smallest number that allows to detect all minima. The number of Earth-distance minima in the full NEODyS database has been computed for increasing values of $N$. For $N\ge48$, all minima -- shown in  Table~\ref{tab:NEOSvsMINIMA} -- are detected. A  more conservative $N=50$ has finally been chosen.

\begin{table}
\centering
\caption{Number of NEOs with a certain number of minimum values of the distance between its orbit and the Earth orbit.}
\label{tab:NEOSvsMINIMA}
\begin{tabular}{rr}
  \hline
  No. Minima & No. NEOs \\
  \hline
  1 & 8186 \\
  2 & 9444 \\
  3 & 18   \\
  4 & 0    \\
  \hline
  Total: & 17648  \\
 Average: & 1.537 \text{min/object} \\
\end{tabular}
\end{table}

The code is able to redo this parametric analysis if the database is updated.

\subsection{Algorithm 2: golden section search iterations}

Algorithm 2 needs to find the value and anomaly of  bracketed local minima. Golden section search has been selected to refine the values. No comparisons have been made, since we consider it clearly superior to other methods, such as bisection and ternary search.

The choice of grid divisions $N$ affects the number of iterations needed to locate the minimum to required precision. A finer grid needs fewer iterations. But $N$ has other requirements: the lowest value that allows to locate all minima.

\begin{table}
\centering
\caption{Golden Section algorithm calls and number of iterations for the 30 first objects of NEODyS database}
\label{tab:GSS_ITERATION}
	\begin{tabular}{rrlrl}
		\hline
		& \multicolumn{2}{c}{Golden Section Search} \\
		No.         & Abs.  & Rel.   \\
		Iterations	& Freq.	& Freq.  \\
		\hline
		28          & 1	    & 0.02128 \\
		29          & 1	    & 0.02128 \\
		30          & 2	    & 0.04255 \\
		31          & 9	    & 0.19148 \\
		32          & 17    & 0.36170 \\
		33          & 5	    & 0.10638 \\
		34          & 5	    & 0.10638 \\
		35          & 2	    & 0.04225 \\
		36          & 4	    & 0.10638 \\
		37          & 1	    & 0.08511 \\
		\hline
		TOTAL :		& 47	& 1.0000 \\
		AVERAGE : 	& \multicolumn{2}{c}{32.467 \text{iter/proc}} \\
	\end{tabular}
\end{table}

Table~\ref{tab:GSS_ITERATION} shows the number of iterations needed to refine all bracketed local minima. This has been done for the first 30 database objects, on the Intel I7 computer aforementioned.

Golden section search has been called 47 times -- as many as minima in the set of 30 NEOs.  The number of  iterations per call ranges from 28 to 37, with an approximate average of 32.5.

\section{Comparison of methods\label{S:COMPARISON}}

\subsection{Gronchi's MOID method\label{SS:GRONCHI}}

Giovanni Gronchi's FORTRAN software for MOID computation of 1995 \citep{Gronchi} has been provided by the author's kindness. It is highly regarded for its speed and precision.

Two computer systems have been used to compare \sdg and Gronchi's methods:
\begin{itemize}
  \item A Supermicro server with 4 Intel Xeon X7560 microprocessors @ \SI{2.27}{\giga\hertz} and \SI{64}{\gibi\byte} of RAM, running Windows Server 2012 OS and provided with Intel C++ and FORTRAN compilers.
  \item A Fujitsu Esprimo personal computer with an Intel I7 microprocessor @ \SI{2.93}{\giga\hertz} and \SI{16}{\gibi\byte} of RAM, running Linux OS and provided with GNU C++ 5.4.0 and GNU FORTRAN 5.4.0 compilers.
\end{itemize}
No parallelization has been implemented, only one core is used on each machine.  Earth MOID has been computed for the 17648 asteroids in the set. Computations have been repeated at different times to discount machine load.

Tables~\ref{tab:FE} and~\ref{tab:SM} show average execution times per MOID for the two computers, each with its set of compilers: C++ for \sdg and FORTRAN for Gronchi's.

\begin{table}
  \centering
  \caption{Average MOID computation times  in the Fujitsu  PC. Run ten times at different moments.}
  \label{tab:FE}
  \begin{tabular}{crr}
     \hline
& \multicolumn{2}{c}{Average unit time} \\
& \multicolumn{2}{c}{\si{\micro\second}} \\
Run &	Gronchi	& \sdg	    \\
\hline
\#1 &	168.43	& 134.16 \\
\#2 &	170.42	& 136.09  \\
\#3 &	167.69	& 135.83  \\
\#4 &	167.55	& 135.14  \\
\#5 &	167.82	& 134.20  \\
\#6 &	167.58	& 134.29  \\
\#7 &	168.25	& 134.49  \\
\#8 &	167.27	& 133.93  \\
\#9 &	166.16	& 135.33  \\
\#10&	168.96	& 136.73  \\
\hline
mean &	168.01	& 135.02 \\
st.dev. &	1.13 &	0.96 \\
\cline{1-2}
\multicolumn{2}{c}{Diff. rel. to Gronchi} & 0.1964	 \\
\hline
Sample size &	15000 &	20$\times$16748	 \\
     \hline
   \end{tabular}
\end{table}

\begin{table}
  \centering
\caption{Average MOID computation times  in the Supermicro server. Run ten times at different moments.}
\label{tab:SM}
  \begin{tabular}{crrr}
     \hline
& \multicolumn{3}{c}{Average unit time}  \\
& \multicolumn{3}{c}{\si{\micro\second}}  \\
Run &	Gronchi	& \multicolumn{2}{c}{\sdg}  \\
 & & Halley & Danby  \\
\hline
\#1   	& 82.78	& 64.72	& 61.18 \\
\#2   	& 77.47	& 64.01	& 68.79 \\
\#3   	& 77.96	& 68.53	& 63.57 \\
\#4   	& 81.19	& 61.62	& 60.47 \\
\#5   	& 81.23	& 68.53	& 67.73 \\
\#6   	& 81.59	& 63.13	& 61.00 \\
\#7   	& 79.24	& 63.48	& 61.54 \\
\#8   	& 77.46	& 61.89	& 60.83 \\
\#9   	& 78.97	& 63.84	& 62.07 \\
\#10  	& 77.51	& 65.69	& 61.09 \\
\hline
Mean    & 79.54 & 64.54	& 62.83	 \\
St. dev.	& 2.00  & 2.42	& 3.00	 \\
\cline{1-2}
\multicolumn{2}{c}{Diff. rel. to Gronchi} & 0.1885	& 0.2101 \\
\hline
Sample size & 20$\times$16748 &	20$\times$16748 & 20$\times$16748 \\
\hline
   \end{tabular}
\end{table}

As for accuracy, the results of \sdg and Gronchi's method are essentially the same. They have been compared asteroid by asteroid, but the spreadsheet would be too large to reproduce. Table~\ref{tab:dif} shows the eleven NEOs with the greatest differences. Even among those, the difference is less than \SI{1.1E-15}{} except for the first entry (2016VB1), which seems to be anomalous. This needs clarification.

Throughout the development of the \sdg method, a reference tool was needed to check the results. The classic two-variable minimisation shown in subsection~\ref{ss:DEE3D} has  been coded in \textsc{Maple}\texttrademark\ 2017.3. Minimisation has been performed over the $[0,2\pi]\times[0,2\pi]$ anomaly domain, with 30-digit precision. \textsc{Maple} routines were used. This method is very accurate, but slow.

To solve the discrepancy, global minimum for 2016VB1 has been computed with \textsc{Maple}. The result is the same as that of  \sdg's method to 16 digits, the latter's accuracy. The complete values, primary and secondary anomalies and distance, are shown in Table~\ref{tab:MINIMA}.

\begin{figure}
  \centering
  \includegraphics[width=.9\columnwidth]{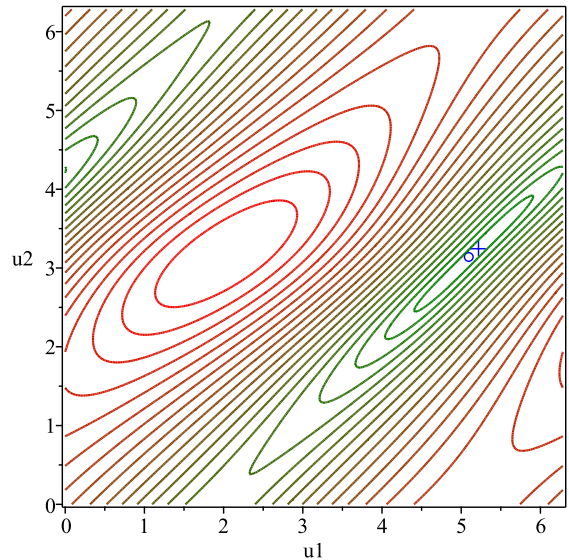}
  \caption{\textsc{Maple} contour plot of distances between the Earth and the asteroid 2016VB1. Red is larger and green smaller. Minimum point symbols: \sdg ($+$), Gronchi ($\circ$).}\label{fig:2016VB1}
\end{figure}

Additionally, a contour map of the distance as a function of anomalies -- as in \citet{Armellin} -- has been generated with \textsc{Maple} (see Fig.~\ref{fig:2016VB1}). This allows to see all extrema and their character. The minima found by \sdg and Gronchi's methods are marked on the plot. These are two very close minima, in a narrow valley, one smaller than the other. It is easy to slip from one to the other.

\begin{table*}
 \centering
 \caption{List of NEOs with the greatest MOID differences between Gronchi's and \sdg methods.}
\label{tab:dif}
\begin{tabular}{llll}
  \hline
\multirow{3}{*}{NEO'S NAME} & \multicolumn{2}{c}{MOID$_{\earth}$} & \multicolumn{1}{c}{$\Delta$MOID$_{\earth}$} \\
            & \multicolumn{2}{c}{\si{\astronomicalunit}} & \multicolumn{1}{c}{\si{\astronomicalunit}} \\
            & \multicolumn{1}{c}{GRONCHI}               & \multicolumn{1}{c}{\sdg}                  & |\sdg-GRONCHI| \\
  \hline
2016VB1	    &0.0019763068271733     &0.0012415773444360    &0.0007347294827373 \\
2014SE      &0.0342838294450500	    &0.0342838294450511    &0.0000000000000011 \\
159560	    &0.4733258977758740		&0.4733258977758730	   &0.0000000000000011 \\
2005QO11	&0.3352818313812930		&0.3352818313812920	   &0.0000000000000011 \\
2005TS45	&0.2892358593469260		&0.2892358593469250	   &0.0000000000000011 \\
2006AV2	    &0.2668343096980130		&0.2668343096980140	   &0.0000000000000011 \\
2010BH2	    &0.3557077619773020		&0.3557077619773030	   &0.0000000000000011 \\
2012BD27	&0.4626880416478720		&0.4626880416478730	   &0.0000000000000011 \\
2012TA79	&0.3278131305485040		&0.3278131305485050	   &0.0000000000000011 \\
2013GY73	&0.3041646227553150		&0.3041646227553160	   &0.0000000000000011 \\
2017FO2	    &0.2802988699612530		&0.2802988699612520	   &0.0000000000000011 \\
  \hline
\end{tabular}
\end{table*}

\begin{table*}
\caption{Two different minima of the distance between the Earth and the asteroid 2016VB1 orbits with the corresponding anomalies.}
\label{tab:MINIMA}
\begin{tabular}{llll}
  \hline
  \multirow{2}{1cm}{Method} & \multicolumn{1}{c}{$u_1$}  & \multicolumn{1}{c}{$u_2$}  & \multicolumn{1}{c}{$d(u_1,u_2)$ \eqref{eq:du1u2}}  \\
  & \multicolumn{1}{c}{\si{\radian}} & \multicolumn{1}{c}{\si{\radian}} & \multicolumn{1}{c}{\si{\astronomicalunit}} \\
  \hline
  Gronchi & 5.1044487302561962 & 3.1308165142757937 & \SI{1.97630682717E-3}{} \\
  \sdg     & 5.2180959925761516 & 3.2444773076892670 & \SI{1.24157734443E-3}{} \\
  \hline
\end{tabular}
\end{table*}

Remarkably, only one out of 17648 cases shows discrepancy between both methods. For all others, differences are equal to or smaller than \SI{1.1E-15}{\astronomicalunit}. Note that both methods use double precision floats, and \sdg's tolerance is $xtol=\;$\SI{4.44E-16}{}.

\subsection{Other MOID methods\label{SS:OTHERS}}

Since we have no access to codes for other methods, direct comparisons are not possible.  Two authors include tests against Gronchi's method, both in speed and accuracy, which allow some indirect conclusions.

\citet{Wisnioski} report average unit times of \SI{400}{\micro\second} per MOID for Gronchi's methods and from \SIrange{90}{150}{\micro\second} for their method. But no indication is given about the equipment with which calculations have been performed.

\citet{Derevyanka} obtains average unit times of \SI{120}{\micro\second}  without activating parallelization, versus \SI{400}{\micro\second} for Gronchi's. They use a PC with Intel Core i7-4800MQ CPU @ \SI{2.7}{\giga\hertz}, \SI{8.0}{\gibi\byte} of DDR3 RAM, and a \SI{1}{\tera\byte} HDD at \SI{7200}{\per\minute}.

The results match Gronchi's up to at least 9 exact figures, so it could be inferred that a tolerance of about \SI{5E-10}{} is being used. Derevyanka's code is not available. If this tolerance is set in \sdg method and a computer similar to Derevyanka's is used -- that mentioned in  subsection~\ref{ss:HvsD} -- average unit times are less than \SI{23.5}{\micro\second}.

We cannot directly compare accuracy, but both authors acknowledge that their speed involves some loss of precision. Since our results match Gronchi's to 15 exact figures, we can conclude that \sdg is more precise than the two mentioned above.

\section{Conclusions\label{SS:CONCL}}

The mathematical justification given in section~\ref{S:MFSDG} makes unnecessary the exhaustive calculation of all critical points of the distance between Keplerian orbits to obtain MOID. The method described in this paper computes MOID faster than the other  methods considered, with a smaller risk of omitting or confusing minima.

Depending on the computer used, this method is  between 19 and 20 per cent faster than that in \citet{Gronchi}, as Tables~\ref{tab:FE} and~\ref{tab:SM} show. It is not a dramatic improvement, but it also gives an equal or superior precision in locating global minima, which is a common weakness of some methods for multivariate optimization.

\section*{Acknowledgements}

This work is part of the project entitled  \textit{Dynamical Analysis of Complex Interplanetary Missions} with reference {\bf ESP2017-87271-P}, supported by Spanish {\bf Agencia Estatal de Investigaci\'on (AEI)}  of {\bf Ministerio de Econom\'ia, Industria y Competitividad (MINECO)}  and by European Found of Regional Development (FEDER).

We are grateful to Prof. Roberto Armellin for his detailed and helpful review of the manuscript.



\bibliographystyle{mnras}
\bibliography{MOID_Hedo_2018} 



\appendix

\section{Properties related to the use of the squared of the objective function\label{APP:PROPERTIES}}

Consider the non-negative function $ F(u)\in \mathrm C^2$.
\begin{alignat}{2}
f(u)  = \frac{\mathrm{d} F^2(u)}{\mathrm d u} &= 2 F(u) F'(u)                   \label{eq:dps} \\
f'(u) =  \frac{\mathrm{d}^2 F^2(u)}{\mathrm d u^2} &= 2( F(u) F''(u) +  F'(u)^2) \label{eq:dss}
\end{alignat}

It is easy to prove the following results:
\begin{enumerate}
\item The roots $u_r$ of a non-negative function $ F(u)\in \textrm{C}^1$ are also stationary points.

  Proof by contradiction\\
  Suppose that $\exists u_r\in\mathbb{R} :  F(u_r)=0\ $ and $\  F'(u_r)\not=0$
  \begin{alignat*}{2}
  &\exists \delta>0 :  F\left(u_r - \delta\, \sign{ F'(u_r)} \right) =\\
  &\text{(Th. of mean value of Taylor's expansion)}  \overset{\exists \gamma\in [0,1]}{\underset{}{=}} \\
  &=\cancel{ F(u_r)} - \delta\, \sign{ F'(u_r)}  F'\left(u_r- \gamma \delta\, \sign{ F'(u_r)}\right) < 0 =\\
  &= F(u_r) \\
  &\text{Contradiction, because $ F(u_r)=0$ must be a global minimum}\\
  &\implies  F'(u_r)=0
  \end{alignat*}

\item The stationary points of $ F(u)$ coincide with those of its square $F(u)^2$.

  Proof\footnote{Numbers in parentheses above equal or imply signs are references to the equations used to deduce the equality.}
  \begin{alignat*}{2}
  &u^*\ \text{stationary point of }  F \implies  F'(u^*)=0 \implies\\
  &\stackrel{\eqref{eq:dps}}{\implies}\ f(u^*)=0 \quad\implies\quad u^*\ \text{stationary point of } F^2\\
  &u^*\ \text{stationary point of } F^2 \implies  f(u^*)=0 \implies\\
  &\stackrel{\eqref{eq:dps}}{\implies}\ \left\{
    \begin{array}{ll}
       F'(u^*)=0\ &:\ u^*\ \text{stationary point of }  F\\
       F(u^*)=0\ &:\ \text{(i)}\implies u^*\ \text{stationary point of }  F
    \end{array}\right.
  \end{alignat*}

\item The character of the stationary points of $ F(u)$ is preserved in $F^2(u)$.

    Proof
    \begin{alignat*}{2}
    &\forall u^* :  F'(u^*)=0  \implies\\
    &\implies \sign{f'(u^*)} \stackrel{\eqref{eq:dss}}{=}\sign{2 F(u^*) F''(u^*)} =\\
    &=\left\{
    \begin{array}{ll}
                       0, & \hbox{if }  F(u^*)=0; \hbox{ (Null minima)}  \\
    \sign{ F''(u^*)}, & \hbox{if }  F(u^*)\not=0; \hbox{ (Other critical points)}
    \end{array}
    \right.
    \end{alignat*}

\end{enumerate}

\section{Cases with closed-form solution for the distance between an ellipse and a coplanar point\label{APP:SOLUTIONS}}

Each case is considered with the two approaches referred to in~\ref{ss:DPE2D}.

\noindent {{Circular case ($a=b=R$)}}.

\begin{enumerate}
\renewcommand{\theenumi}{Approach (\roman{enumi})}
\item The hyperbola in \eqref{eq:RIH} degenerates into a diametrical line. Equations \eqref{eq:RIH}, \eqref{eq:RIE} and \eqref{eq:DPEC} give
\begin{alignat}{2}
y &\stackrel{\eqref{eq:RIH}}{=} \frac\beta\alpha x   \implies\\
\implies (x,y) &\stackrel{\eqref{eq:RIE}}{=} (\pm \frac{R\alpha}{\sqrt{\alpha^2+\beta^2}},\ \pm \frac{R\beta}{\sqrt{\alpha^2+\beta^2}}) \implies \\
\implies d_{min} &\stackrel{\eqref{eq:DPEC}}{=} |\sqrt{\alpha^2+\beta^2}-R|
\end{alignat}

\item It is a trivial minimization problem with a closed-form solution
\begin{alignat}{2}
u_* &\stackrel{\eqref{eq:FDEU}}{=} \arctan\left(\frac{\beta}{\alpha}\right) \implies\\
\implies d_{min} (u_*;R,R,\alpha,\beta) &\stackrel{\eqref{eq:DPEP}}{=}  |\sqrt{\alpha^2+\beta^2}-R|
\end{alignat}
\end{enumerate}

\noindent {{Elliptical case ($a>b$)}}.

\noindent Subcase $\alpha=0, \beta=0$ ($P\equiv O$)
    \begin{enumerate}
\renewcommand{\theenumi}{Approach (\roman{enumi}):}
    \item The hyperbola degenerates into its asymptotes. Equations \eqref{eq:RIH}, \eqref{eq:RIE}, and \eqref{eq:DPEC} give
    \begin{flalign}
    xy &\stackrel{\eqref{eq:RIH}}{=}0  \implies\\
    \implies (x^E,y^E) &\stackrel{\eqref{eq:RIE}}{=}
    \left\{\begin{array}{l}
        (\pm a,0), \hbox{always} \implies d_{max} \stackrel{\eqref{eq:DPEC}}{=} a\\
        (0,\pm b), \hbox{always} \implies d_{min} \stackrel{\eqref{eq:DPEC}}{=} b
    \end{array}\right.&
    \end{flalign}

    \item The roots of $f(u)$ are $u^*\in \{0,\frac{\pi}{2}, \pi,-\frac{\pi}{2}\}$. Both roots $u_*=\pm\frac{\pi}{2}$ give the same minimum distance $d(\pm\frac{\pi}{2}) = b$, because $f'(\pm\frac{\pi}{2})>0$ (relative minima) and $f'(0)<0, f'(\pi)<0$  (relative maxima). It is easy to see that $d(0)=d(\pi)=a > b = d(\pm\frac{\pi}{2})$.
    \end{enumerate}

\noindent Subcase $\alpha=0, \beta>0$ ($P$ in the positive side of $Oy$)
  \begin{enumerate}
    \renewcommand{\theenumi}{Approach (\roman{enumi}):}
    \item The hyperbola degenerates into its asymptotes. Equations \eqref{eq:RIH}, \eqref{eq:RIE}, and \eqref{eq:DPEC} give
    \begin{flalign}
    &x\left(y(a^2-b^2)+b^2\beta\right) \stackrel{\eqref{eq:RIH}}{=}0 \quad\implies \\
    &(x^E,y^E)
    \stackrel{\eqref{eq:RIE}}{=}
    \left\{\begin{array}{lr}
        (0,\pm b),\ &\rightarrow \\
        \left(\pm\sqrt{1-\frac{\beta^2 (1-e^2)}{a^2 e^4}},-\frac{\beta(1-e^2)}{e^2}\right),\  &\rightarrow
    \end{array}\right. \nonumber\\
    &\left.\begin{array}{l}
        \rightarrow \hbox{always} \implies  d_{min} \stackrel{\eqref{eq:DPEC}}{=} |\beta-b| \\
        \rightarrow \hbox{if } \beta\le \frac{a e^2}{\sqrt{1-e^2}} \implies d_{max} \stackrel{\eqref{eq:DPEC}}{=} \sqrt{1-e^2}\sqrt{a^2-\frac{\alpha^2}{e^2}}
      \end{array}\right\}
    \end{flalign}

    \item The roots of $f(u)$ are at least $u^*\in\{-\frac{\pi}{2},\frac{\pi}{2}\}$, but $u^*=\arcsin(-\frac{b\beta}{c^2})$ also could be a root if $b\beta\le c^2$. This condition is equivalent to $\beta\le \frac{ae^2}{\sqrt{1-e^2}}$, which means that $P$ is on or inside the plane evolute of $\mathfrak{e}$. The root $u_*=\frac{\pi}{2}$ gives minimum distance $d(\frac{\pi}{2}) = |\beta-b|$, because $f'(\frac{\pi}{2})>0$ (relative minimum). It is easy to see that the relative minimum in $-\frac{\pi}{2}$ when $P$ is on or inside the plane evolute of $\mathfrak{e}$ satisfies $d(-\frac{\pi}{2}) = \beta+b > |\beta-b| =  d(\frac{\pi}{2})$ and that the distance to other stationary points is even greater.
  \end{enumerate}

\noindent Subcase $\alpha>0, \beta=0$ ($P$ in the positive side of $Ox$):
  \begin{enumerate}
    \renewcommand{\theenumi}{Approach (\roman{enumi}):}
    \item The hyperbola degenerates into its asymptotes. Equations \eqref{eq:RIH},\eqref{eq:RIE}, and \eqref{eq:DPEC} give:
    \begin{flalign}
    &y\left(x(a^2-b^2)-a^2\alpha\right) \stackrel{\eqref{eq:RIH}}{=}0 \quad\implies \\
    &(x^E,y^E)
    \stackrel{\eqref{eq:RIE}}{=}
    \left\{\begin{array}{lr}
    (\pm a,0),                                                                       & \rightarrow \\
    \left(\frac\alpha{e^2},\pm a\sqrt{1-e^2}\sqrt{1-\frac{\alpha^2}{e^2a^2}}\right), & \rightarrow
    \end{array}\right. \nonumber\\
    &\left.\begin{array}{ll}
    \rightarrow\ \hbox{always;}             &\rightarrow\ d \stackrel{\eqref{eq:DPEC}}{=} |\alpha-a| \\
    \rightarrow\ \hbox{if } \alpha\le e^2 a &\rightarrow\ d \stackrel{\eqref{eq:DPEC}}{=} \sqrt{(1-e^2)(a^2-\frac{\alpha^2}{e^2})}
      \end{array}
    \right\}\\
    &\text{if }\alpha> e^2 a \implies d_{min} = |\alpha-a|\\
    &\text{if }\alpha\le e^2 a \implies d_{min} = \sqrt{(1-e^2)(a^2-\frac{\alpha^2}{e^2})}
    \end{flalign}

    \item the roots of $f(u)$ are at least $u^*\in\{0,\pi\}$, but $u^*=\arccos(\frac{a\alpha}{c^2})$ also could be a root if $a\alpha\le c^2$. This condition is equivalent to $\alpha\le a e^2$, which means that $P$ is on or inside the plane evolute of $\mathfrak{e}$. Two cases can be distinguished:
    \begin{itemize}
      \item If $\alpha> a e^2$ the root $u_*=0 $ gives the minimum distance $d(0) = |\alpha-a|$, because $f'(0)<0$ (relative minimum) and $f'(\pi)>0$ (relative maximum). It is easy to see that $d(\pi)=\alpha+a > |\alpha-a|= d(0)$.
      \item If $\alpha\le a e^2$  both roots $u_*=\arccos(\frac{a\alpha}{c^2})$ give the minimum distance $d\left(\arccos(\frac{a\alpha}{c^2})\right) = \sqrt{(1-e^2)(a^2-\frac{\alpha^2}{e^2})}$ because $f'(\arccos(\frac{a\alpha}{c^2}))= a^2e^2-\frac{\alpha^2}{e^2}>0$ (relative minimum) and $f'(0)\le 0$ (possible relative maximum or inflection with horizontal tangent). It is easy to see that $d(0) = |\alpha-a| \ge \sqrt{(1-e^2)(a^2-\frac{\alpha^2}{e^2})} = d(\arccos(\frac{a\alpha}{c^2}))$.
    \end{itemize}
  \end{enumerate}

\section{Halley's method\label{APP:HALLEY}}

The Halley's method, or hyperbolic approximation (\cite{Wesstein}) is an iterative root-search procedure for  class $C^2$ functions. It can be considered simultaneously as:
\begin{itemize}
   \item A Householder-type method of second order (with cubic convergence);
   \item A Newton--Raphson method applied to the function $g(x) = \dfrac{f(x)}{\sqrt{f'(x)}}$.
\end{itemize}
It has been chosen for solving the root of $f(u)$ because:
\begin{enumerate}
  \item It is simple to implement, with explicit expressions for all derivatives.
  \item It has fewer problems with double roots.
  \item It can recover from stagnation when a stationary point is reached under certain conditions.
  \item It has cubic convergence against the quadratic one of  Newton--Raphson.
\end{enumerate}

The increment of iteration is defined by
\begin{equation}
\Delta(x) = - \frac{f(x) f'(x)}{f'(x)^2-\frac{1}{2}f(x)f''(x)} \label{eq:H2}
\end{equation}

\begin{itemize}
\item Algorithm progression condition
\begin{equation}
    \Delta(x) \not=0 \implies f(x) f'(x) \not=  0
\end{equation}
Cases:
\begin{enumerate}
  \item If $f(x^*)=0 \implies x^*$ is the desired root. Exit.
  \item If $f'(x^*)=0$ the method stagnates, as in Newton--Raphson methods.
\end{enumerate}

\item Algorithm convergence conditions
\begin{alignat}{2}
h_0 &= \Delta(x_0)= - \frac{g(x_0)}{g'(x_0)} \\
J_0 &=
\left\{
  \begin{array}{ll}
    ( x_0 , x_0 + h_0 ), & \text{if } h_0 > 0 \\
    ( x_0 + h_0 , x_0 ), & \text{if } h_0 < 0 \\
  \end{array}
\right. \\
\forall x &\in J_0,\  \exists M_0 > 0 : |g''(x)| < M_0, |g'(x)| > 2 |h_0| M_0
\end{alignat}

\item Halley's method stagnates at the stationary points of the function ($x^* \in \mathbb{R} : f '(x^*) = 0$). A quick exit is needed. The Householder's method of order 3 (the next order of Halley's without this drawback) can be used
\begin{equation}
\Delta(x) = - \frac{6 f(x) f'(x)^2 - 3 f(x)^2 f''(x)}{6 f'(x)^3-6 f(x)f'(x)f''(x) +f(x)^2 f'''(x)}
\end{equation}
which near stationary points becomes
\begin{equation}
\Delta(x) \stackrel{f'(x)=0}{=} \frac{3 f''(x)}{f'''(x)} \label{eq:H3}
\end{equation}

Therefore, if $f''(x) \not = 0$ (avoiding stagnation) and $f'''(x) \not = 0$ (avoiding the singularity), we break out by substituting $\Delta (x)$ of \eqref{eq:H3} for that of \eqref{eq:H2}. We must also ensure that the new value falls within the convergence zone of the searched root, which is governed by other convergence conditions.
\end{itemize}


\bsp	
\label{lastpage}

\end{document}